# Disordered hyperuniformity signals functioning and resilience of self-organized vegetation patterns

Wensi Hu [1], Quan-Xing Liu [2,*], Bo Wang [1], Nuo Xu [1], Lijuan Cui [3,*], Chi Xu [1,4,*]

[1] School of Life Sciences, Nanjing University, Nanjing 210023, China;

[2] School of Mathematical Sciences, Shanghai Jiao Tong University, Shanghai 200240, China;

[3] Chinese Academy of Forestry, Beijing Key Laboratory of Wetland Services and Restoration, Beijing, China;

[4] Breeding Base for State Key Laboratory of Land Degradation and Ecological Restoration in northwestern China; Key Laboratory of Restoration and Reconstruction of Degraded Ecosystems in northwestern China of Ministry of Education, Ningxia University, Yinchuan 750021, China.

[*] Authors for correspondence: Quan-Xing Liu (qx.liu@sjtu.edu.cn), Lijuan Cui (lkyclj@126.com), and Chi Xu (xuchi@nju.edu.cn)

**Abstract**

In harsh environments, organisms may self-organize into spatially patterned systems in various ways. So far, studies of ecosystem spatial self-organization have primarily focused on apparent orders reflected by regular patterns. However, self-organized ecosystems may also have cryptic orders that can be unveiled only through certain quantitative analyses. Here we show that disordered hyperuniformity as a striking class of hidden orders can exist in spatially self-organized vegetation landscapes. By analyzing the high-resolution remotely sensed images across the American drylands, we demonstrate that it is not uncommon to find disordered hyperuniform vegetation states characterized by suppressed density fluctuations at long range. Such long-range hyperuniformity has been documented in a wide range of microscopic systems. Our finding contributes to expanding this domain to accommodate natural landscape ecological systems. We use theoretical modeling to propose that disordered hyperuniform vegetation patterning can arise from three generalized mechanisms prevalent in dryland ecosystems, including (1) critical absorbing states driven by an ecological legacy effect, (2) scale-dependent feedbacks driven by plant-plant facilitation and competition, and (3) density-dependent aggregation driven by plant-sediment feedbacks. Our modeling results also show that disordered hyperuniform patterns can help ecosystems cope with arid conditions with enhanced functioning of soil moisture acquisition. However, this advantage may come at the cost of slower recovery of ecosystem structure upon perturbations. Our work highlights that disordered hyperuniformity as a distinguishable but underexplored ecosystem self-organization state merits systematic studies to better understand its underlying mechanisms, functioning, and resilience.

**Introduction**

Orders may arise spontaneously from self-organization in a wide range of spatially extended natural systems, such as sand dunes (*1, 2*), coastal mussel beds (*3, 4*), and dryland vegetation (*5-7*). The defining features of such ordered systems are regular shapes or spatially periodic patterns woven across the landscapes, often referred to as 'regular patterns' (see (*5*) for examples). While regular patterns are often easily distinguished in the field or from remotely-sensed images, signaling the existence of spatial self-organization, there are cases that the self-organized orders cannot be easily identified through visual checking, but rather can only be unveiled through analyzing certain quantitative properties. An example of such cryptic orders is that dryland or coastal patchy vegetation may present non-regular, seemingly disordered, patterns, but their patch-size distributions could be well characterized by power laws (or alike) arising from scale-dependent feedbacks (*8-11*). So far, studies on spatial self-organization have primarily focused on apparent orders characterized by regular patterns. Unveiling hidden spatial orders therefore has the potential to expand the self-organization theory into the domain of previously thought disordered systems.

Here we show that 'disordered hyperuniformity' as a striking class of hidden orders can exist in spatially self-organized vegetation landscapes. The term 'disordered hyperuniformity' has been mostly used in physics literature to describe a special geometric property defined by a scale-dependent feature: at small length (spatial) scales, the system presents isotropic structure, resembling typically disordered systems (in contrast with ordered lattice structures); while at large scales, density fluctuations of the system components are suppressed, presenting lower variations than expected by chance, as typically found in crystal lattice structures (*12, 13*). Such orders are considered 'hidden' because they are long-ranged, thus can hardly be perceived intuitively (*13*). In recent years, disordered hyperuniform states have been found in a wide range of systems, starting to invoke much interest across multiple disciplinary fields for studying both non-life and life systems (*14-22*). Interestingly, disordered hyperuniformity can also arise as a result of self-organization in microscopic biological systems such as avian photoreceptor cells (*20*), immune systems (*19*), and microbial populations (*21*), fulfilling important biological functions. However, so far it remains unclear if this distinguishable state may exist in macroscopic ecological systems such as vegetated landscapes.

By analyzing the high-resolution remotely sensed images across the western US, we demonstrate that it is not uncommon to find disordered hyperuniform dryland vegetation patterns spanning across hundreds of meters. We further propose three spatial self-organization mechanisms potentially underlying disordered hyperuniformity of dryland vegetation. We then use mathematical models to examine these mechanisms and probe if the disordered hyperuniformity could signal ecosystem functioning and resilience in water-limited environments.

**Conventional spatial point pattern analysis**

Considering that disordered hyperuniformity has been mostly discovered in point-based spatial structures (*13*), here we focus on spot-like dryland vegetation patterns that are akin to such many-point systems. We used the remotely sensed images available from Google Earth at 0.3 m spatial resolution to arbitrarily select 50 sites with seemingly homogeneous spot-like vegetation patterns across the arid/hyper-arid regions in the western US. For each study site, we extracted the vegetation patches within a plot sized of ~500×500 m$^2$ for subsequent analyses (Fig. 1; see Methods).

As a first step, we conducted spatial point-pattern analysis as a conventional approach for quantifying vegetation patterns (*23, 24*). We calculated the spatial statistics of pair correlation $g(r)$ function (fig. S1) and Ripley's $L$ function (fig. S2) based on vegetation patch centroids (Methods). Our results show a clear signal of over-dispersion typically within the spatial scale of ~10 m (about 2-5 times the mean vegetation patch diameter of 2-4 m); while at larger scales >10 m, the vegetation patterns resemble those generated by Poisson processes in 2D space, often referred to as 'complete spatial randomness (CSR)' in ecology literature (Fig. 2A, D). These features were consistently found across all study sites (some sites also present a signal of under-dispersion at intermediate scales; figs. S1, S2). Therefore, from this conventional approach one may reach the conclusion that the studied dryland vegetation ubiquitously presents 'small-scale over-dispersion and large-scale randomness'. Indeed, such scale-dependent patterning feature has been repeatedly documented in a wide range of dryland ecosystems, and has often been interpreted as a sign of strong plant-plant competition for soil water (resulting in small-scale over-dispersion) (*6, 7, 25*).

**Detecting disordered hyperuniformity**

Despite the similarity of vegetation patterning suggested by the $g(r)$ and Ripley's $L$ statistics, a key between-site difference can be revealed by analyzing density fluctuations. Density fluctuations as a common method for quantifying spatial structure across spatial scales have been mostly applied to microscopic systems (*26-28*). This method quantifies the variance of the system component density $\sigma_\rho^2(L)$ within a window sized of $\sim L^R$ in R-dimensional space, yielding the scaling relation $\sigma_\rho^2(L) \sim L^{-\lambda}$. In this sense, density fluctuations clearly differ from the distance-based spatial point pattern statistics such as the $g(r)$ function.

Our density fluctuations analysis reveals that around 90% of the sites present the scaling exponent λ of density fluctuations <2.0 (mean = 1.14; s.d. = 0.35) consistently across the analysis scales within ~150 m (Fig. 2B). In contrast, 8 study sites present substantially larger values of λ > 2.0 (mean = 2.30; s.d. = 0.18) at long range of ~20-150 m (Fig. 2E). Poisson point processes in 2D space can generate spatial structures with λ = 2.0, therefore the exponent value 2.0 is normally taken as a key reference of Poisson patterns (*13*). Apparent deviations of λ from 2.0 towards larger values are seen as a clear indication of long-range hyperuniformity; more specifically, the situation that λ falls between 2 and 3 (as found in our study) is referred to as Class III disordered hyperuniformity (*13*). An intuitive interpretation is that the decay of density fluctuations is faster than the increase of window size ($\sim L^2$, the exponent 2.0 is the spatial dimension value; 2D Poisson patterns thus present density fluctuations exponent λ equal to 2.0), indicating the suppression of long-range density fluctuations (*13*) To further test the robustness of the observed signal of long-range hyperuniformity, in each site we also analyzed 5 arbitrary plots next to the pre-selected study plot. We found that 6 of the 8 study sites consistently presented long-range hyperuniformity across all plots (fig. S5). These 6 sites are therefore considered to exhibit a robust signal of disordered hyperuniformity (Figs. 1 and 2).

Theoretically, an equivalent way to test for long-range hyperuniformity is to look at the power spectrum characterized by the structure factor $S(k)$ (*12*). Specifically, when $k$ (a variable indicating wave number) approaches 0, or in other words, spatial scale becomes infinitely large, disordered hyperuniformity presents $S(k) \to 0$, typically characterized by a scaling relation $S(k) \sim k^\alpha$ with the exponent $\alpha > 0$ (see (*12, 13*) for details). Indeed, our results showed that this criterion is commonly met, reinforcing the existence of long-range hyperuniformity in these study sites (Fig. 2F).

Our results thus clearly demonstrate that disordered hyperuniformity as a distinguishable state can exist in dryland vegetation. If disordered hyperuniformity is not rare in the real-world ecosystems (in a rough sense, $>\sim 10\%$ of the somewhat arbitrarily selected landscapes across a broad geographic extent in our study), one may speculate that such long-range spatial patterning could be underpinned by some common ecological processes, raising the essential question: what are the generalized mechanisms, if any, underlying disordered hyperuniformity of dryland vegetation?

**Potential mechanisms**

Unraveling this question is clearly a difficult task, requiring systematic studies combining long-term observations, field experiments, and theoretical modeling. Here we hypothesize that disordered hyperuniform vegetation patterning can arise from three mechanisms including (1) critical absorbing states driven by an ecological legacy effect, (2) scale-dependent feedbacks driven by plant-plant facilitation and competition, and (3) density-dependent aggregation driven by plant-sediment feedbacks. These mechanisms are not meant to be exhaustive, nor to specifically address the analyzed American dryland systems. Instead, they are meant to serve as a starting point towards a more comprehensive and generalized mechanistic understanding of landscape-scale disordered hyperuniformity. We elaborate on these hypotheses and quantitatively examine them using three mathematical models in the following sections.

*Critical absorbing states*

Recent physical studies have shown that 'critical absorbing states' of many-particle systems can give rise to disordered hyperuniformity (*18, 29, 30*). Briefly put, a system consisted of many movable particles can present a phase transition between an active phase characterized by never-ending dynamics of the particles and an absorbing phase in which the particles cease to move; and disordered hyperuniformity can arise at the critical (density) point of this phrase transition with gradually increasing particle density (*29*). Importantly, due to its simplicity and robustness, it is reasonable to speculate that critical absorbing states may potentially act as a generalized mechanism underlying disordered hyperuniformity across a range of systems.

At first glance, it seems unlikely that vegetation patterns pertain to movement-driven critical absorbing states, as sessile plants cannot literally move like active matter (e.g. (*21*)). However, vegetation systems may also present similar re-organization behavior that resembles particle movement, when taking into account the mortality-regeneration process: assuming that the system has a constant population density at a certain carrying capacity, the mortality of an individual will then allow the successful establishment of a newly generated individual; if the newborns have different spatial locations than the deceased ones, the system is spatially re-organized in a manner that mimics particle movement. Moreover, it has been shown that small random displacements close to critical absorbing states can lead to robust disordered hyperuniformity subject to, for instance, noises or periodical perturbations (*18*). Thinking in parallel, for dryland vegetation systems, this mechanism could be realized simply by an ecological legacy effect that makes the newborns tend to establish in close proximity to the deads (in a general sense, ecological legacy can be defined as the pronounced influences of previously existing biotic or abiotic conditions on current ecological properties (*31-33*)). A common effect in dryland ecosystems is that established plants can substantially increase the survival probability of newly generated plants nearby through ameliorating e.g. soil water and nutrient conditions (known as the 'nurse plant effect', (*33-35*)). Such facilitation may persist even after the mortality of benefactors, for instance, the soils can maintain higher water infiltration rates for years to facilitate seedling regeneration and survival (*36, 37*), creating a positive legacy effect from the dead plants.

We use individual-based modeling to test if this ecological legacy effect may give rise to disordered hyperuniformity. As a reference, we start with a model assuming that competition between neighboring plants is the only non-neutral process in the modeled system ('competition model' hereafter), considering that strong water competition often acts as a dominant biotic interaction between plants. In this reference model, random birth and death of plants take place periodically in a 2D space. The survival probability of any given newborn depends on its distance to existing plants, i.e., survival probability decreases dramatically when the newborn falls within a certain range of the competitive interaction ($R_c$) to the existing ones (see Methods for details). Plant competition is short-ranged, thus reasonably cannot lead to long-range hyperuniformity - this can be confirmed by our results showing that at large scales the exponent λ of the density fluctuations scaling relation falls around 2.0 (Fig. 3D), and the exponent α of the structural factor scaling relation falls around 0 (Fig. 3F).

We then test if the inclusion of the ecological legacy effect can suppress long-range density fluctuations. We build upon the competition model by assuming that a newborn is randomly located in proximity to deceased individuals in a manner inverse to the competition effect ('ecological legacy model' hereafter; see Method). Our simulation results revealed that the ecological legacy model can generate long-range hyperuniformity characterized by a scaling relation of density fluctuations $\sigma_\rho^2(L) \sim L^{-2.45}$ and a structure factor $S(k) \sim k^{0.45}$. Our further analysis confirms that the disordered hyperuniformity indeed arises around the critical point of the transition between the active phase and absorbing phase (*29, 30*) (fig. S6). At the critical points, the ecological legacy model gives rise to systems with higher plant densities than the counterparts resulting from the competition model with the same parameter settings, suggesting that the ecological legacy mechanism facilitates higher levels of abundance (Fig.

3E).

*Scale-dependent feedbacks*

Extensive studies have shown that scale-dependent feedbacks (i.e., short-range positive feedback and long-range negative feedback) following the Turing's activation-inhibition principle are widespread, and often play a key role in shaping patchy vegetation patterns in drylands (e.g. (*5, 6, 25, 38*)). A common class of scale-dependent feedbacks hinges on plant-plant interactions: established plants can provide crucial facilitation to their closely co-existing neighbors (for instance the aforementioned nurse plant effect), while interfere the plants at longer distances through soil water competition (for instance realized by lateral roots that extend far beyond the canopy extent) (*25*).

We build upon the aforementioned competition model by adding the facilitation effect to mimic the scale-dependent feedbacks driven by plant-plant interactions ('scale-dependent feedbacks model' hereafter). Therefore, the only difference between the scale-dependent feedbacks model and the ecological legacy model is that the facilitation is produced by alive vs. dead plants. Our parameterized scale-dependent feedbacks model can give rise to spot-like 'Turing patterns'. Importantly, these modeled vegetation patterns clearly present long-range hyperuniformity, as indicated by the density fluctuations scaling relation $\sigma_\rho^2(L) \sim L^{-2.5}$ and structure factor $S(k) \sim k^{0.5}$. It should be noted that such long-range hyperuniformity can only arise in certain parameter settings. For example, the scaling exponent λ gradually decreases with elevating environmental harshness (a parameter used for controlling the background mortality) from 2.5 to 2.0, indicating the loss of long-range hyperuniformity.

*Density-dependent aggregation*

Very recent work highlights that 'density-dependent aggregation' following the phase separation principle in physics can potentially act as a generalized mechanism giving rise to patchy patterns across a range of ecological systems (including dryland vegetation) that apparently lack scale-dependent feedbacks (*39*). Compared with scale-dependent feedbacks, density-dependent aggregation has major differences in system dynamics (see (*39*) for details). However, at some particular stages during the aggregation processes the resulting spatial patterns may appear more or less similar to those from scale-dependent feedbacks, in terms of the emergence of (quasi) spatial periodicity (*39*).

It has been speculated that the (quasi) periodic patterns resulting from density-dependent aggregation could follow disordered hyperuniformity (*39*), but rigorous tests remain lacking. Here we analyzed the density-dependent aggregation model with respect to sediment trapping proposed by (ref (*39*)). This model ('density-dependent aggregation model' hereafter) mimics the essential plant-sediment feedbacks underlying nebkhas-like biomorphic systems with hummocked structures typically found across the global drylands (*40-42*): established plants can enhance deposition rates and/or decreasing erosion rates of aeolian sands, and the trapped sediments can in turn facilitate the plants through increasing soil water and nutrient availability, and help the plants escape adverse conditions due to erosion and elevated soil salinity (*41, 42*). This density-dependent aggregation model can also give rise to long-range hyperuniformity, as indicated by the density fluctuations scaling relation $\sigma_\rho^2(L) \sim L^{-3.0}$ and structure factor $S(k) \sim k^{4.0}$ (Fig. 4L). Moreover, the arising disordered hyperuniformity is not restricted to spot-like patterns but also pertains to patterns mimicking labyrinths or gaps (Fig. 4H, I). Our results thus suggest that disordered hyperuniformity may come into being as a transient state during the phase separation processes.

**Disordered hyperuniformity may enhance ecosystem functioning**

Why would ecosystems spatially self-organize into disordered hyperuniform states? There is increasing suggestion that spatial self-organization can enhance key aspects of ecosystem functioning (*5, 7, 43*). A corollary is that disordered hyperuniform states can do the same job. We test this idea using a simple moisture process model (see Methods), by assessing if and to what extent hyperuniform and non-hyperuniform patterns differ in the acquisition of soil moisture, which is arguably among the most critical aspect for dryland ecosystems (*44-46*).

We assess this ecosystem function as the moisture fraction of plant uptake relative to loss through soil evaporation. In our moisture process model, soil moisture comes from a rainfall event with a fixed amount and spatially homogeneous distribution; subsequently the soil moisture is reduced through evaporation at a rate $r_e$, and in the meantime depleted by plant uptake at a rate $r_p$, until all available soil moisture is exhausted. The soil moisture available for each plant individual comes from the spatial domain identified by the Voronoi algorithm (the landscape is partitioned into polygons, each of which represents the root domain of an individual plant). We assess the soil moisture acquisition function as the fraction of plant uptake relative to loss through soil evaporation.

We compare four types of spatial patterns with equal plant densities: (1) ordered hyperuniform pattern represented by a hexagonal lattice (characterized by a density fluctuation scaling relation $\sigma_\rho^2(L) \sim L^{-3}$), (2) disordered hyperuniform pattern generated by the ecological legacy model (i.e., critical absorbing states, characterized by $\sigma_\rho^2(L) \sim L^{-2.45}$), (3) disordered non-hyperuniform pattern with small-scale over-dispersion generated by the competition model (characterized by $\sigma_\rho^2(L) \sim L^{-2}$), and (4) disordered non-hyperuniform pattern with CSR (characterized by $\sigma_\rho^2(L) \sim L^{-2}$). Note that here we focus on the functioning of these 'static' patterns (see Fig. 5A), thus the patterns are held unchanged without considering pattern formation mechanisms.

We use the performance of the hexagonal lattice pattern as the reference, which presents the highest function of soil moisture acquisition (Fig. 5B, C). The disordered hyperuniform pattern shows a slightly lower function level than the hexagonal lattice pattern, but can substantially outperform the non-hyperuniform patterns (Fig. 5B, C). Such advantage of the disordered hyperuniformity tends to be more pronounced at higher evaporation rates (Fig. 5B) and lower plant uptake rates (Fig. 5C). This finding aligns with the intuition that if the resource is homogeneously distributed in space, more uniform distributions of plant individuals can lead to more uniform partitioning of the space, in turn, faster system-wide uptake, leaving less moisture evaporated. Indeed, it has been shown that disordered hyperuniformity may pertain to 'jammed particle packing' in the sense that disordered hyperuniformity tends to present denser packing (*47*).

**Resilience of disordered hyperuniform patterns**

Another speculation is that disordered hyperuniform states may enhance the resilience of dryland vegetation to drought events, as positive links between ecosystem resilience and self-organized ecosystem patterns have been documented in stressful environments (*48, 49*). Using perturbation experiments in modeled ecosystems, we quantify the recovery rate as an engineering resilience indicator to test this speculation.

We compare the hyperuniform patterns generated by the ecological legacy (critical absorbing states) model and the scale-dependent feedbacks model to the non-hyperuniform pattern generated by the competition model, to probe how they could respond to drought-induced mortalities. We do not include the hexagonal lattice pattern and the CSR pattern considering that these patterns lack clear pattern formation mechanisms as a necessary ingredient of recovery.

Interestingly, our results demonstrate that after perturbation the non-hyperuniform pattern recovers faster than the hyperuniform patterns underlain by both scale-dependent feedbacks (Fig. 6A) and ecological legacy effect (Fig. 6B). A close scrutinizing to the models reveals that in stochastic morality-regeneration processes, it takes longer for the system to get settled in the hyperuniform states representing more 'packed' spatial configuration. On the other hand, when the system is purely driven by local competition, the convergence to equilibrium is faster. Therefore, if stochastic re-organization of the systems is a key element for pattern formation, hyperuniform states may come at the cost of slower recovery of system spatial structure upon perturbations. Such spatial aspect of resilience could be essentially different from the previously studied resilience upon spatially homogeneous perturbation without invoking spatial structure re-organization (*48, 50-52*).

It should be noted that we did not take into account density-dependent aggregation as this model does not represent typical equilibria, making it difficult to compare recovery rate.

However, previous work has suggested that systems driven by density-dependent aggregation tend to have lower resilience than those driven by scale-dependent feedbacks (*39*). Assessing the resilience of these patterns driven by specific formation mechanisms is an intriguing topic but requires in-depth theoretical and empirical studies.

**Outlook and summary**

Our results demonstrate that dryland vegetation can self-organize into disordered hyperuniform states characterized by suppressed density fluctuations at long range. Such hidden order of spatial self-organization is increasingly documented in a wide range of microscopic physical, chemical, and biological systems. Our work contributes to expanding this domain to include natural landscape ecological systems that are largely underexplored. More importantly, our results suggest that disordered hyperuniform vegetation states are plausibly underlain by three generalized mechanisms prevalent in dryland ecosystems. Our findings invite future studies to systematically scan vegetation patterns and test pattern formation mechanisms from the lens of disordered hyperuniformity. Our work also links disordered hyperuniform patterns to ecosystem functioning and resilience, facilitating the understanding of how ecosystems may adapt to stressful environments and respond to climate change by self-organizing into disordered hyperuniform states.

While there is still a long way to go toward unraveling the mechanisms underlying disordered hyperuniform vegetation states, previous studies on microscopic systems have gained valuable insights (*21*). The known mechanisms are of various forms, but many pertain to long-range interactions. For instance, algae with circularly swimming behavior may generate particular fluid flow that results in effective repulsions between the algae individuals in the long range. Such long-range hydrodynamic interactions can suppress density fluctuations and drive the algae to self-organize into disordered hyperuniform states (*21*). However, long-range interactions are not a necessary condition - critical absorbing states purely driven by local interactions can also give rise to disordered hyperuniformity (*29, 30*). Indeed, it has been shown in our models that both local (ecological legacy effect and density-dependent aggregation) and non-local (scale-dependent feedbacks) interactions are possible mechanisms for dryland vegetation systems. Nonetheless, these theoretical predictions need to be tested and refined with detailed field observations and thoughtful experiments. For example, it is essential to understand on what conditions long-range hyperuniformity can arise in systems governed by Turing or phase separation principle.

While the underlying mechanisms could commonly occur in real-world ecosystems, implying that disordered hyperuniform patterns may be frequently found in nature, their existence may be restricted by a range of potential limiting factors. Our modeling analyses suggest that the existence of disordered hyperuniformity could be sensitive to a range of factors such as environmental conditions, biotic/abiotic interactions, and ontogeny of pattern formation. Moreover, environmental heterogeneity (such as soil patchiness, and topographic heterogeneity) may easily diminish the signal of long-range hyperuniformity. It is worth noting that we found 22 sites presenting hyperuniform signals (the density fluctuations exponent >2.0) within the intermediate scale range (around 10-40 m) but disappearing at larger scales (fig. S7). If such lack of hyperuniformity signals at sufficiently long ranges can be attributed to large-scale environmental heterogeneity, the frequency of inherent disordered hyperuniform patterns is likely even higher. While this class of intermediated-range density fluctuation suppression is not considered to strictly satisfy the criterion of disordered hyperuniformity, it may also carry useful information to better understand pattern formation, and thus may merit further studies. Third, our results suggest that the formation of disordered hyperuniform patterns may take a long time as reflected by their relatively low structural resilience, therefore these patterns are likely sensitive to frequent environmental perturbations that disrupt their formation. Taken together, we need both in-depth theoretical analyses and broad-scale empirical searching (*53*) for disordered hyperuniform patterns to understand the genesis of this distinctive ecosystem state.

Mounting evidence converges to suggest that spatial self-organization may help the organisms cope with stressful environments by enhancing system functioning and/or resilience. However, our results point to the possibility that essential tradeoffs may exist between ecosystem functioning and ecosystem (structural) resilience. An implication is that the link between self-organization and ecosystem functioning/resilience may be more complex and context dependent than previously thought. Echoing G. E. Hutchinson's classic formulation 'ecological theater and evolutionary play' (*54*), the perspective of 'survival of the systems' (*55*) may help to further unravel this link towards a better understanding of spatial self-organization.

**Materials and Methods**

*Data collection and Image analyses*

We scanned the North American drylands using the global aridity index dataset (areas falling into the categories of 'hyper-arid', 'arid', and 'semi-arid' (*56*), Fig. 1A) in search for (seemingly) homogeneous spot-like vegetation patterns. We focused on the areas where high-resolution (0.3 m) remotely sensed images are available from Google Earth. We avoided areas with apparent landscape heterogeneity with respect to topographic features and human land use. The search was conducted visually using Google Earth in an arbitrary way, from which 50 sites were finally included in our study (Fig. 1B, see table. S1 for details). Within each site we set up a plot sized of ~500×500 m$^2$. For each plot, we georegistered the Google Earth images and then extracted vegetation patches using a supervised classification approach with a maximum likelihood method. We then converted the vegetation patches into polygons and extracted their centroids for subsequent analyses. The image analyses were conducted in QGIS 3.18 and MATLAB 2021a.

*Conventional spatial point pattern analyses*

We used the vegetation patch centroids to conduct spatial point pattern analyses based on the Ripley's *L* function and pair correlation $g(r)$ function (*57*). These conventional spatial statistics have been widely used for quantifying vegetation patterns in terms of over- or under-dispersion at multiple scales. For instance, the $g(r)$ function, as a summary statistic, is a measure of the probability of finding two individuals at a given distance (*r*) from each other.

    If $g(r)$ derived from the focal vegetation pattern is significantly higher than that derived from the null model characterizing random distributions (normally Poisson point process, also termed as complete spatial randomness, CSR), the vegetation pattern is interpreted as 'under-dispersion', and as 'over-dispersion' if lower than the null model. Ripley's *L* function is a similar spatial statistic but has been suggested to be less accurate than the $g(r)$ function (*57*). We show the results from the $g(r)$ function only in the main text, and the results from the Ripley's *L* function are highly consistent with those from the $g(r)$

function (fig. S2). We used R version 4.2.1 (*58*) with the 'spatstat' package (*24*) to conduct these spatial point pattern analyses. See ref (*57*) for details on the spatial point pattern analyses.

*Density fluctuations and structure factor*

The density fluctuations method essentially quantifies how the density of system components varies across spatial scales. For 2D systems, density fluctuations are calculated as the variance of density $\sigma_\rho^2(L) \equiv \langle \rho^2(L) \rangle - \langle \rho \rangle^2$ within sampling windows with a fixed size $L$ randomly laid across the system. Varying window size $L$, the systems following Poisson distributions can yield a scaling relation $\sigma_\rho^2 \sim L^{-2}$. When density fluctuations decay faster than Poisson distributions, the scaling relation yields $\sigma_\rho^2 \sim L^{-\lambda}$ with $2 < \lambda \leq 3$. This indicates that component distribution presents lower variations (or in other words, more 'uniform') than 'random distributions', an indicator of the presence of hyperuniformity (*13*).

Previous studies mostly pertain to microscopic systems with relatively even component sizes (such as particles or cells), therefore can reasonably treat the components as points. However, vegetation landscapes present considerably higher variations in the sizes of plant individuals or vegetation patches (see fig. S8 for vegetation patch size distributions for the study sites). We focus on quantifying mass density fluctuations in line with previous work (*59-61*). Unfortunately, it is impossible to accurately retrieve biomass distributions using Google Earth images, because the near-infrared band data are not available and the available images have been transformed in histograms in unknown ways. We therefore used $(S_i)^{\frac{3}{2}}$ as a surrogate variable for the biomass of a given vegetation patch $i$ ($S_i$ denotes the patch area; the exponent 3/2 is due to the assumption that biomass is proportional to 3D volumn; for instance, for a spherically shaped object with the area $S_i$ projected on a 2D surface, its 3D volume is $\frac{4}{3\sqrt{\pi}}(S_i)^{\frac{3}{2}}$). $\rho(L)$ is thus calculated as:

$$\rho(L) = \frac{4}{3\sqrt{\pi}L^2}\sum_{i=1}^{n}(S_i)^{\frac{3}{2}} \tag{1}$$

where $n$ denotes the number of patches within a sampling window sized of $L$.

The structure factor $S(k)$ describes the distribution of point density in systems over a range of wavelength, $L$:

$$S(k) = \frac{1}{N} \langle |\sum_{j=1}^{N} e^{-i\boldsymbol{k}\cdot\boldsymbol{r}^{(j)}}|^2 \rangle \tag{2}$$

where $N$ is the total number of points, $\boldsymbol{r}$ is the particle position, $\boldsymbol{k}$ is the wave number, $k = 2\pi/L$. This method is therefore essential based on the characterization of power spectral density.

Details of the density fluctuations and structural factor methods can be found in (*13*).

### Simple models

#### The competition model

We built a simple individual-based model to account for the effect of plant-plant competition on vegetation patterning adapted from (*62*). In the competition model, each plant individual is represented by a point in a continuous 2D space. The competition model starts from randomly generating $n$ points; if the distance between two plant individuals is smaller than a predefined threshold $R_c$ (i.e. 'hardcore'), one of the individual (selected randomly) dies. Following the mortality of $m$ deceased individuals, $m$ new individuals are generated randomly, the total individual nubmer thus is maintained at $n$. Such a mortality-regeneration process is repeated until no individual dies. We also consider a 'softcore' version, where the between-plant competition is modeled by a Hill function as in (*10*). Both the hardcore and softcore versions obtain consistent patterns (fig. S9). See table S2 for model parameters.

#### The ecological legacy model

The ecological legacy effect model was built by adapting the regeneration process in the competition model. Instead of generating new individuals randomly, now new individuals are generated in proximity to the dead individuals to mimic the facilitative effect of the previously existing plants. Specifically, the newborns are distributed randomly within the interaction range of $R_c$ from the deceased ones. When the system is stabilized (no individual dies), 20% of individuals are randomly selected and die. In this way, the system can spatially re-organize to achieve hyperuniform states that are robust to perturbations around the critical density point (*18*) (fig. S10).

*The scale-dependent feedbacks model*

We built a minimal scale-dependent feedbacks model by adapting (*10*). In this model, individuals are randomly generated followed by mortality. The mortality probability of a given offspring individual $i$ ($M_i$) is a function of environmental harshness ($h_e$) and the distance ($d_j$) to all individuals $j$:

$$M_i = \sum_{j=1}^{all} \left(\text{competition}_j - \text{facilitation}_j\right) + h_e \qquad (3)$$

The facilitation and competition strengths are modeled as (sigmoidal) Hill functions of distance to individual $j$:

$$facilitation_j = \frac{h_f^{p_f}}{d_j + h_f^{p_f}} \qquad (4)$$

$$competition_j = \frac{\alpha * h_c^{p_c}}{d_j + h_c^{p_c}} \qquad (5)$$

where the exponents $p_f$ and $p_c$ determine the steepness of the sigmoids, and $h_c$ and $h_f$ are the half-saturation constants. Short-range facilitation and long-range competition are modeled by setting $h_f$ large than $h_c$. In addition, $p_f$ is set twice as large as $p_c$ to account for the observation that facilitation usually vanishes more quickly than competition with distance.

The scale-dependent feedbacks model as well as the aforementioned competition model and the ecological legacy model are individual-based models. We conduct the density fluctuation analyses using the spatial points (equally sized) from the model outputs.

*The density-dependent aggregation model*

We adopted the sediment trapping model by (*39*) to consider the density-dependent aggregation process. Within this system, sediments are in a deposited or a mobile state (due to wind erosion). Established vegetation can increase the deposition rate and reduce the erosion rate (sediment trapping). The sediments trapped by plants can benefit plant growth by enhancing soil moisture and nutrient availability. Additionally, it aids the plants in coping with adverse conditions arising from erosion and elevated soil salinity levels. The model is expressed as follows:

$$\frac{\partial S_d}{\partial t} = D(P)S_s - \frac{\alpha}{\beta+P(S_d)} + D_d\left(\frac{\partial^2 S_d}{\partial x^2} + \frac{\partial^2 S_d}{\partial y^2}\right) \quad (6)$$

$$\frac{\partial S_s}{\partial t} = \tau\left(\frac{\alpha}{\beta+P(S_d)} - D(P)S_s\right) + D_s\left(\frac{\partial^2 S_s}{\partial x^2} + \frac{\partial^2 S_s}{\partial y^2}\right) \quad (7)$$

$$D(P) = \varepsilon P(S_d) \quad (8)$$

$$P(S_d) = \xi S_d \quad (9)$$

where $S_s$ represents the concentration of suspended/aeolian sediments, $S_d$ is the height of plant-trapped sediment and $\tau$ converts $S_d$ into $S_s$. The deposition rate $D(P)$, which indicates the rate at which suspended/aeolian sediment is trapped, is assumed to be proportional to the plant biomass $P(S_d)$. Plant biomass is assumed to increase linearly with the height of deposited sediment $S_d$. The constants involved in the model are $\alpha, \beta, \xi,$ and $\varepsilon$. See (*39*) for the model details.

Due to the inherent complexity, the density-dependent aggregation model is built based on partial differential equations. The model can produce vegetation biomass distribution maps as output (as shown in Fig. 4G-I). We therefore can directly conduct the analyses of density fluctuations based on biomass distribution without invoking the weight of vegetation patches.

*Soil moisture acquisition function*

We used a simple moisture model to assess the efficiency of soil moisture acquisition by vegetation following a rainfall event (with total available water amount $P$). The system presents spatially homogeneous soil evaporation at the rate $r_e = \frac{K(P_s - P_a)}{t}$ (where $K$ is a constant, and $P_s$ is the saturation vapor pressure depedent on temperature). In arid conditions, $P_s$ and $P_a$ are assumed to remain constant due to low vapor pressure and high air circulation, leading to constant $r_e$ in the model.

We use the Voronoi algorithm to partition the space with spatial point (representing individual plants) patterns into distinct non-overlapping polygons, each of which represents the exclusive spatial domain for the plant individual (at the polygon centroid) to use the soil moisture therein ($P_i$, proportional to the polygon area). After time $t_i$, the soil moisture of a given polygonal area is exhausted due to evaporation at a constant rate $r_e$ and plant uptake at a constant rate $r_p$:

$$P_i = (r_e + r_p) \times t_i. \tag{10}$$

Ultimately, the fraction of soil moisture acquisition by vegetation can be calculated as:

$$E = \frac{\sum_{i=1}^{n}(r_p \times t_i)}{P} \tag{11}$$

***Recovery of vegetation patterns upon perturbation***

We assessed the recovery rate upon perturbations as an engineering resilience indicator to compare between disordered hyperuniform patterns and non-hyperuniform patterns. The hyperuniform patterns are generated by the ecological legacy (critical absorbing states) model and the scale-dependent feedbacks model. We consider these two models because they can produce spatial points systems with comparable parameter settings; also, the underlying ecological processes with clear equilibria in these models allow for scruitinzing the recovery processes. We separately compared these two models with their non-hyperuniform counterparts. Specifically, for the ecological legacy model, the non-hyperuniform counterpart is the competition model (Fig. 6B); whereas for the scale-dependent feedbacks model, the non-hyperuniform counterpart is obtained by removing the facilitation component in Eq. (3) (Fig. 6A). Therefore, although these two non-hyperuniform systems as recovery references are both governed by competition, their recovery trajectories are different due to slightly different model settings.

We conducted a perturbation experiment by randomly eliminate 20% plant individuals (points) from the spatial patterns at same plant densities generated by the models. This elimination mimics drought-induced mortalities. We calculated the time steps the systems need to recover to the original plant densities. Note that we focus on the resilience of the spatial patterns, therefore the perturbations destruct the spatial patterns. Such perturbations are different than the studies reducing plant biomass homogeneously across the space.

**Acknowledgements**

This work was supported by the National Key R&D Program of China (2022YFF1301000), the National Natural Science Foundation of China (grant no. 32071609, and 32061143014) and the Open Fund for Key Laboratory of Land Degradation and Ecological Restoration in northwestern China of Ningxia University (grant no. LDER2023Z01).

**Conflict of Interest**

After the completion of this work (first submission on November 13, 2023), we noticed that an observation of disordered hyperuniformity of Turing vegetation patterns was published on October 10, 2023 (doi: 10.1073/pnas.2306514120), however, we do not agree with part of the analyses and results of that published paper.

**Author contributions**

Q.-X.L., L.C., and C.X. designed research. W.H., Q.-X.L. and C.X. developed the models. W.H. carried out image analyses and performed simulations. W.H., Q.-X.L., and C.X. wrote the first draft. All coauthors interpreted results and revised the manuscript.

**Data and materials availability**

All data needed to evaluate the conclusions in the paper are present in the paper and/or the Supplementary Materials. The code used in this study is publicly available at Github after acceptance.

**Figures**

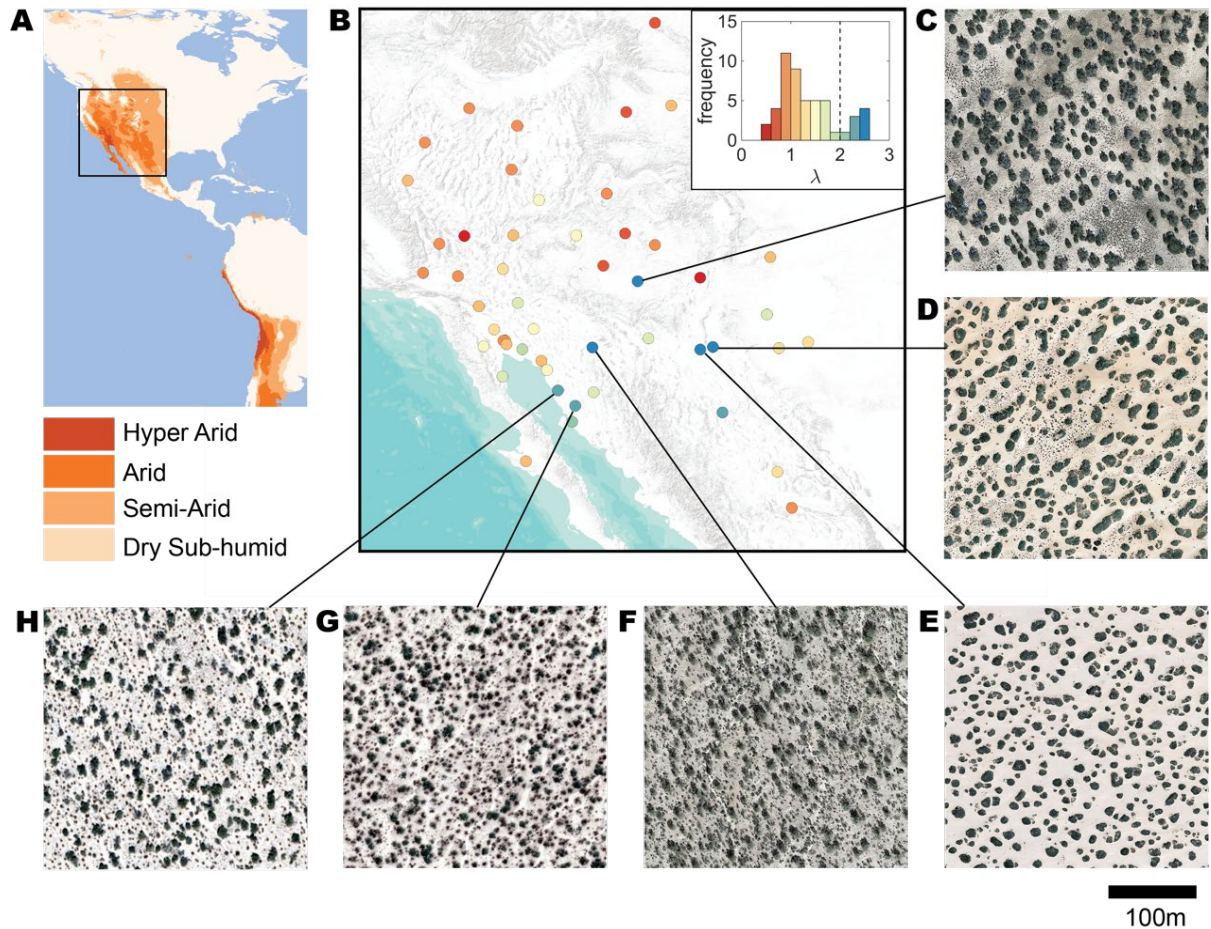

**Figure 1. Locations of all 50 study sites and the high-resolution satellite images of the sites presenting robust signals of disordered hyperuniformity.** **(A)** Geographic extent of the drylands in the western US. **(B)** Locations of the 50 study sites, where the exponent λ of density fluctuations scaling relation ranges from 0.5 to 2.6 (from red to blue). The inset panel shows the frequency distribution of λ for the 50 sites; the dashed line marks λ at 2.0; λ>2.0 signals disordered hyperuniformity. **(C-H)** The high-resolution satellite images available from Google Earth illustrate the typical vegetation spatial patterns (dark representing vegetation patches) presenting robust disordered hyperuniformity signals in six study sites.

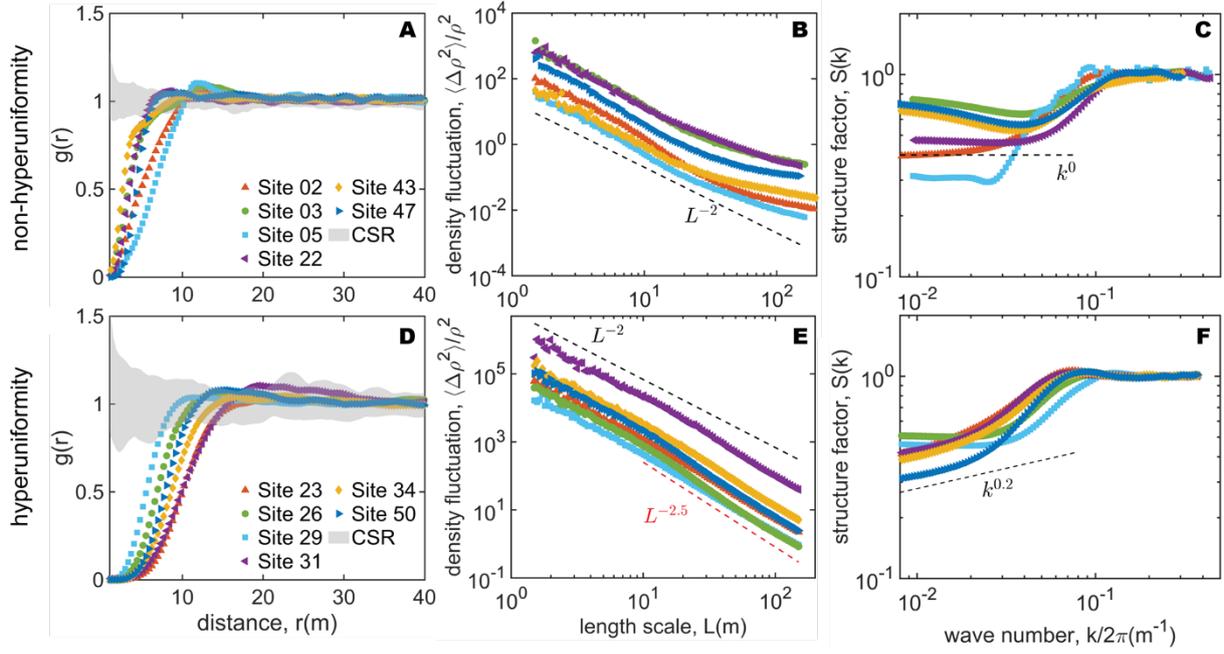

**Figure 2. Comparison between the non-hyperuniform and hyperuniform vegetation patterns. (A-C)** Analyses of non-hyperuniform patterns from 6 typical study sites. See figs. S1-4 for other sites presenting non-hyperuniform patterns. **(D-F)** Analyses of disordered hyperuniform patterns robustly found in the 6 study sites illustrated in Fig. 1. Results from the pair correlation function $g(r)$ show a similar signal of small-scale overdispersion and large-scale randomness; the gray bands represent 95% confidence intervals of complete spatial randomness (CSR) (**A** and **D**). In contrast, density fluctuations analysis shows an important difference characterized by the scaling exponent λ<2.0 **(B)** vs. λ=~2.5 at spatial scales >10 m **(E)**. The exponent values λ larger than 2.0 is an indication of long-range hyperuniformity (see Mthods). The dashed lines represents the scaling relations with exponent λ=2.0 and λ=2.5 as references (**B** and **E**). The existence of long-range hyperuniformity is reinforced by the static structure factor analysis showing $S(k) \sim k^{\alpha>0}$ when $k \to 0$; the dashed lines represents α=0.2 **(F)**.

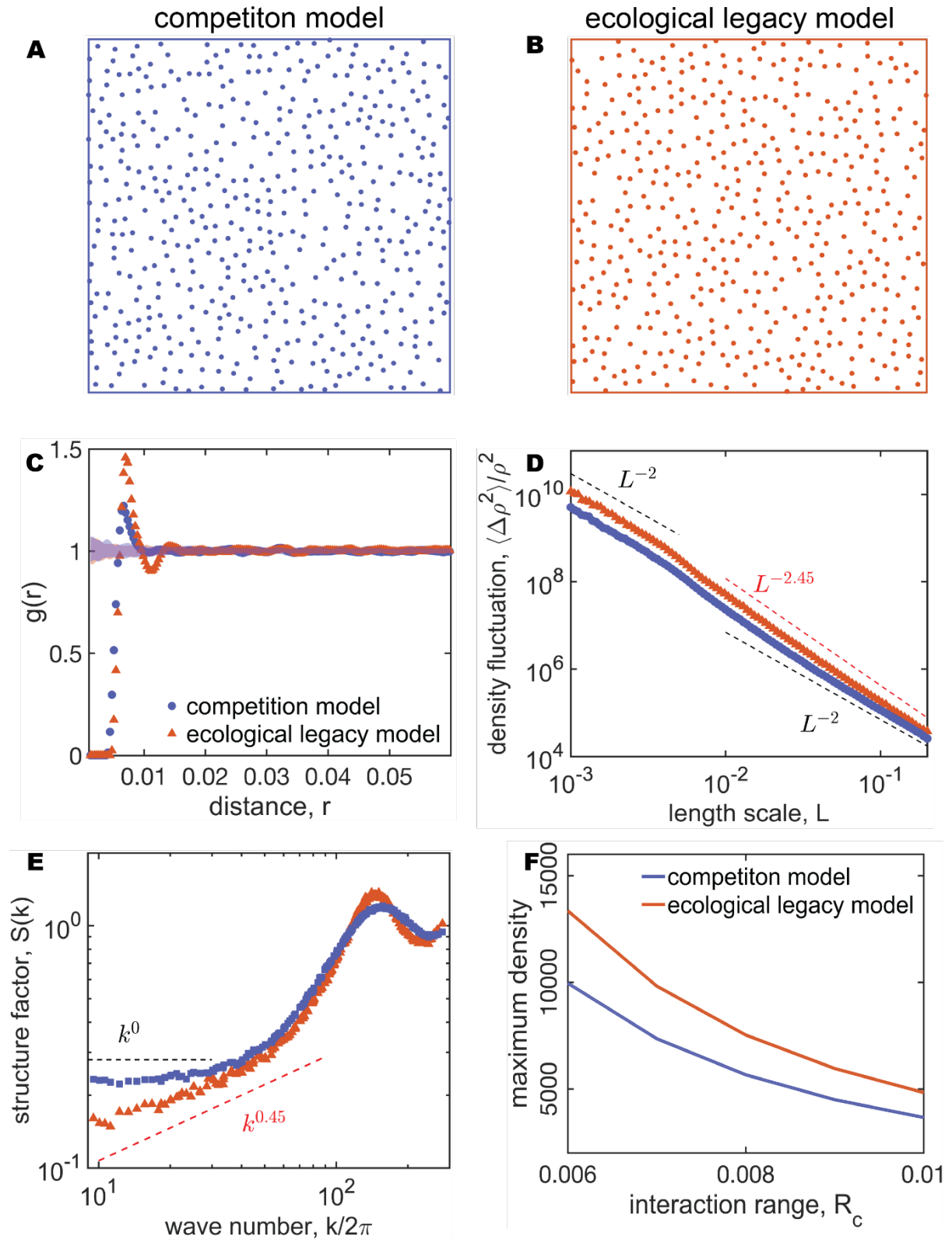

**Figure 3. Comparison between the non-hyperuniform pattern from the competition model and the disordered hyperuniform pattern from the ecological legacy model. (A-B)** Spatial distributions of plant individuals. **(C)** Pair correlation function $g(r)$ show similar

results in terms of small-scale overdispersion and large-scale randomness; the band represents 95% confidence interval of complete spatial randomness. **(D)** The ecological legacy model shows a scaling relation of density fluctuations with exponent λ=2.45, while the competition model shows λ=2.0 at long range. **(E)** The ecological legacy model shows structure factor $S(k) \sim k^{\alpha>0}$ when $k \to 0$, whereas the competition model shows $S(k) \sim k^{\alpha=0}$ when $k \to 0$. **(F)** The ecological legacy model presents higher plant densities than the competition model at equilibria.

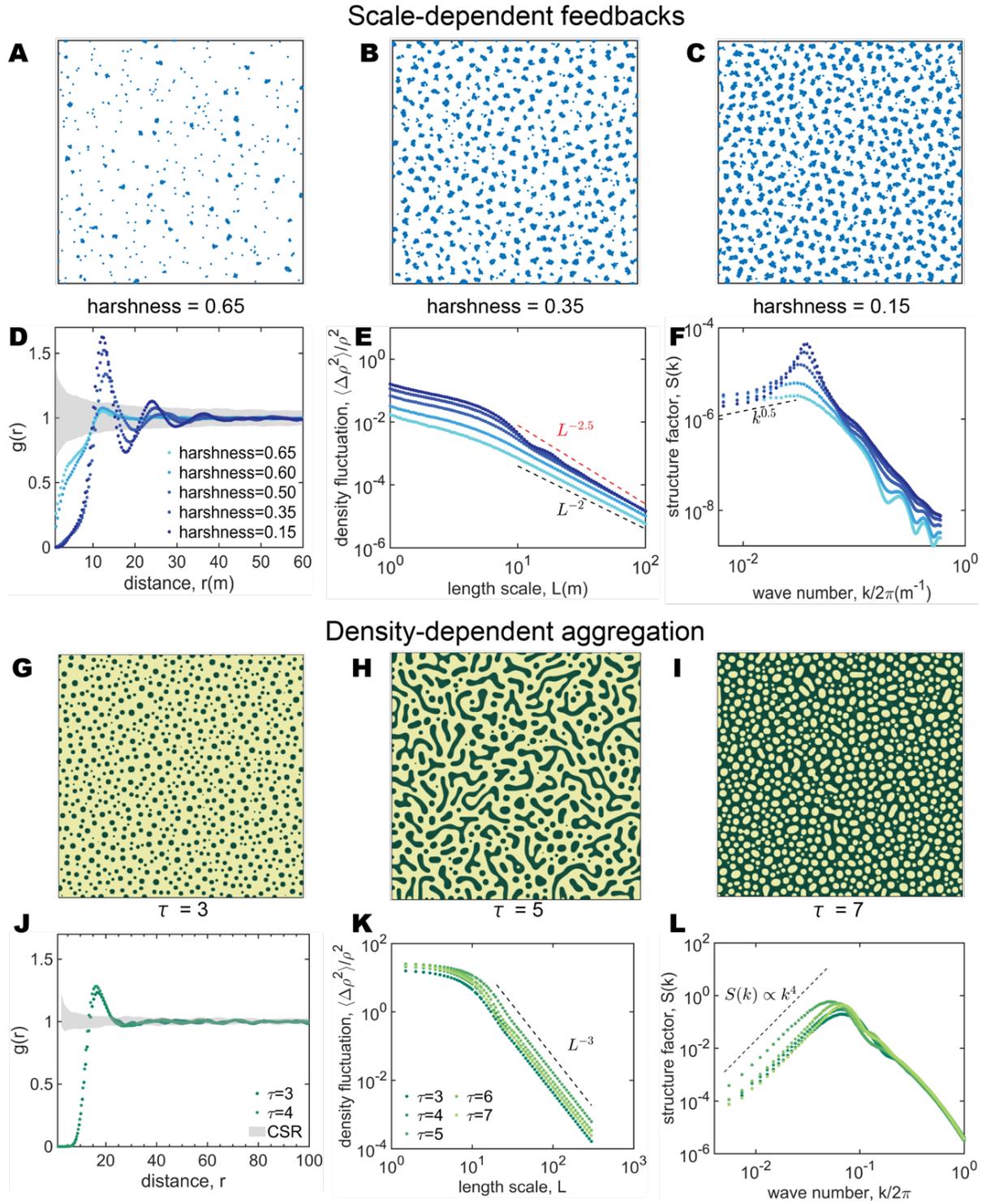

**Figure 4. The disordered hyperuniform patterns from the scale-dependent feedbacks model and the density-dependent aggregation model.** **(A-C)** Spatial distributions of plant individuals from the scale-dpendent feedbacks model with the parameter representing environmental harshness varying from 0.65 to 0.15. **(G-I)** spatial distributions of vegetation patches from the sdensity-dependent aggregation model with changing parameter $\tau$ from 3 to

7 (used to control the level of overall biomass). See Methods for model details. (**D** and **J**) Results from the pair correlation function $g(r)$. (**E** and **K**) The scaling relation of density fluctuations shows exponent λ>2.0 indicating the existence of disordered hyperuniformity. (**F** and **L**) The structure factor shows $S(k) \sim k^{\alpha>0}$ when k→0, also signaling disordered hyperuniformity. Note that the vegetation patterns from the sdensity-dependent aggregation model mimic labyrinths and gaps at larger $\tau \geq 5$, thus the pair correlation function $g(r)$ is not calculated for such non-point systems.

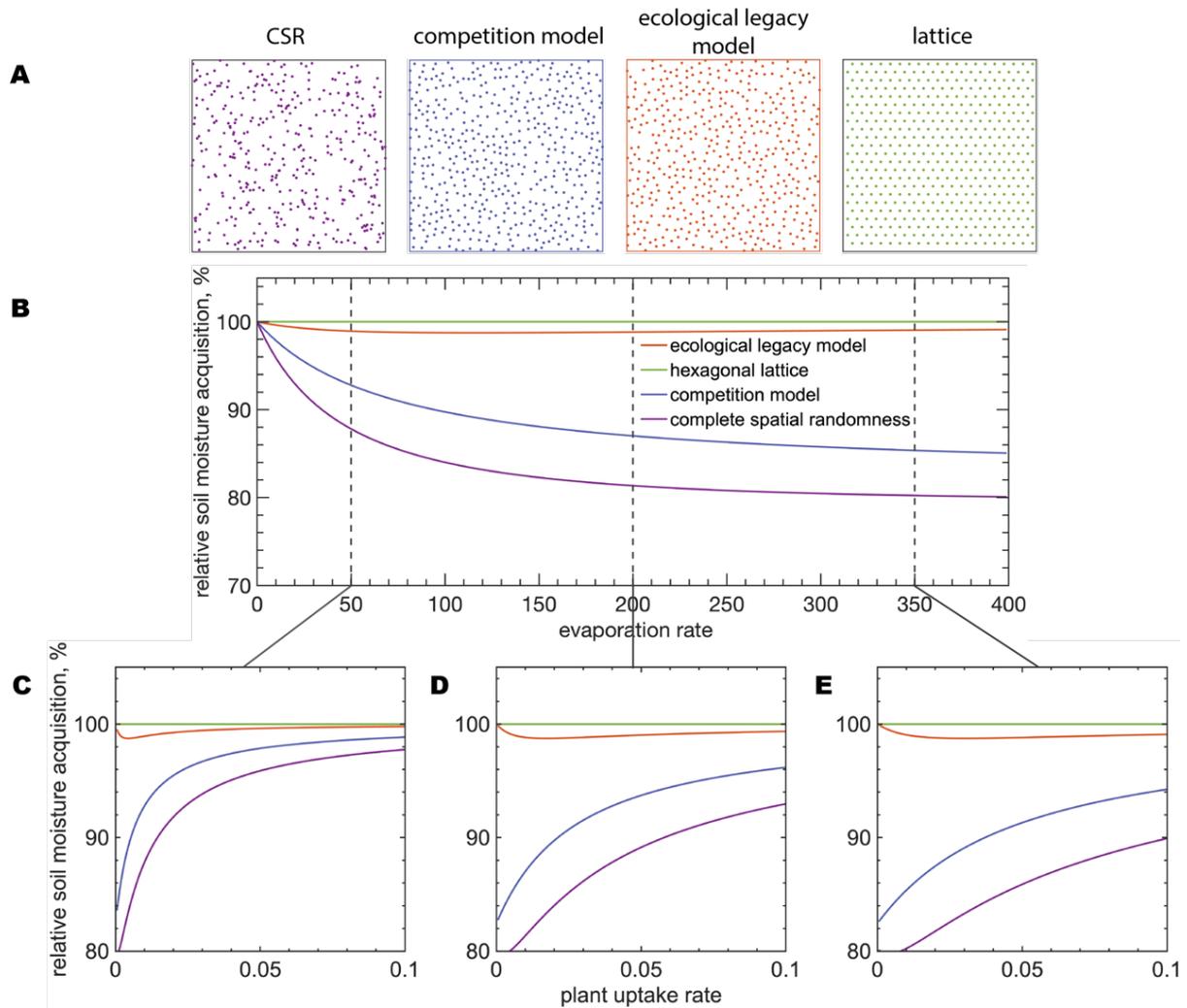

**Figure 5. Assessing the soil moisture acquisition function of four different types of vegetation patterns from a simple moisture process model. (A)** The spatial distributions of plant individuals for the patterns resulting from complete spatial randomness, the competition model, the ecological legacy model, and a hexagonal lattice. **(B)** Relative soil moisture acquisition as a function of evaporation rate, with a fixed plant uptake rate of 0.01. **(C-E)** Relative soil moisture acquisition as a function of plant uptake rate, with fixed evaporation rate of 50 (**C**), 200 (**D**) and 350 (**E**). Note that the hexagonal lattice pattern (green) is taken as a reference, therefore its soil moisture acquisition is held at 100% relative to those of other patterns.

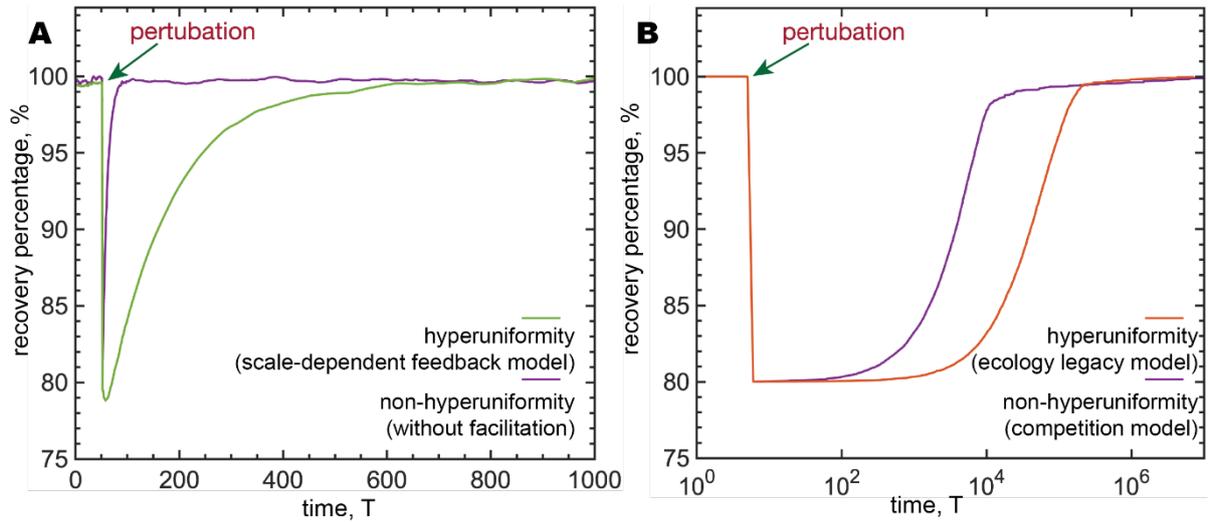

**Figure 6. Post-perturbation recovery of disordered hyperuniform patterns in comparision to non-hyperuniform patterns.** The non-hyperuniform patterns (underlain by only competition) recover faster than the disordered hyperuniform patterns. **(A)** For the disordered hyperuniform pattern derived from the scale-dependent feedbacks model, the non-hyperuniform counterpart is obtained by removing the facilitation component in Eq. (3). **(B)** For the disordered hyperuniform pattern derived from the ecological legacy model, the non-hyperuniform counterpart is the competition model as described in Methods. Although these two non-hyperuniform systems as recovery references are both governed by competition, their recovery trajectories are different due to slightly different model settings. The pertubations are conducted by eliminating 20% plant individuals from the systems at same densities. See Methods for model details.

Supplementary Materials

for

**Disordered hyperuniformity signals functioning and resilience of self-organized vegetation patterns**

Wensi Hu [1], Quan-Xing Liu [2,*], Bo Wang [1], Nuo Xu [1], Lijuan Cui [3, *], Chi Xu [1, 4, *]

[1] School of Life Sciences, Nanjing University, Nanjing 210023, China;

[2] School of Mathematical Sciences, Shanghai Jiao Tong University, Shanghai 200240, China;

[3] Chinese Academy of Forestry, Beijing Key Laboratory of Wetland Services and Restoration, Beijing, China;

[4] Breeding Base for State Key Laboratory of Land Degradation and Ecological Restoration in northwestern China; Key Laboratory of Restoration and Reconstruction of Degraded Ecosystems in northwestern China of Ministry of Education, Ningxia University, Yinchuan 750021, China.

[*]Authors for correspondence: Quan-Xing Liu (qx.liu@sjtu.edu.cn), Lijuan Cui (lkyclj@126.com), and Chi Xu (xuchi@nju.edu.cn)

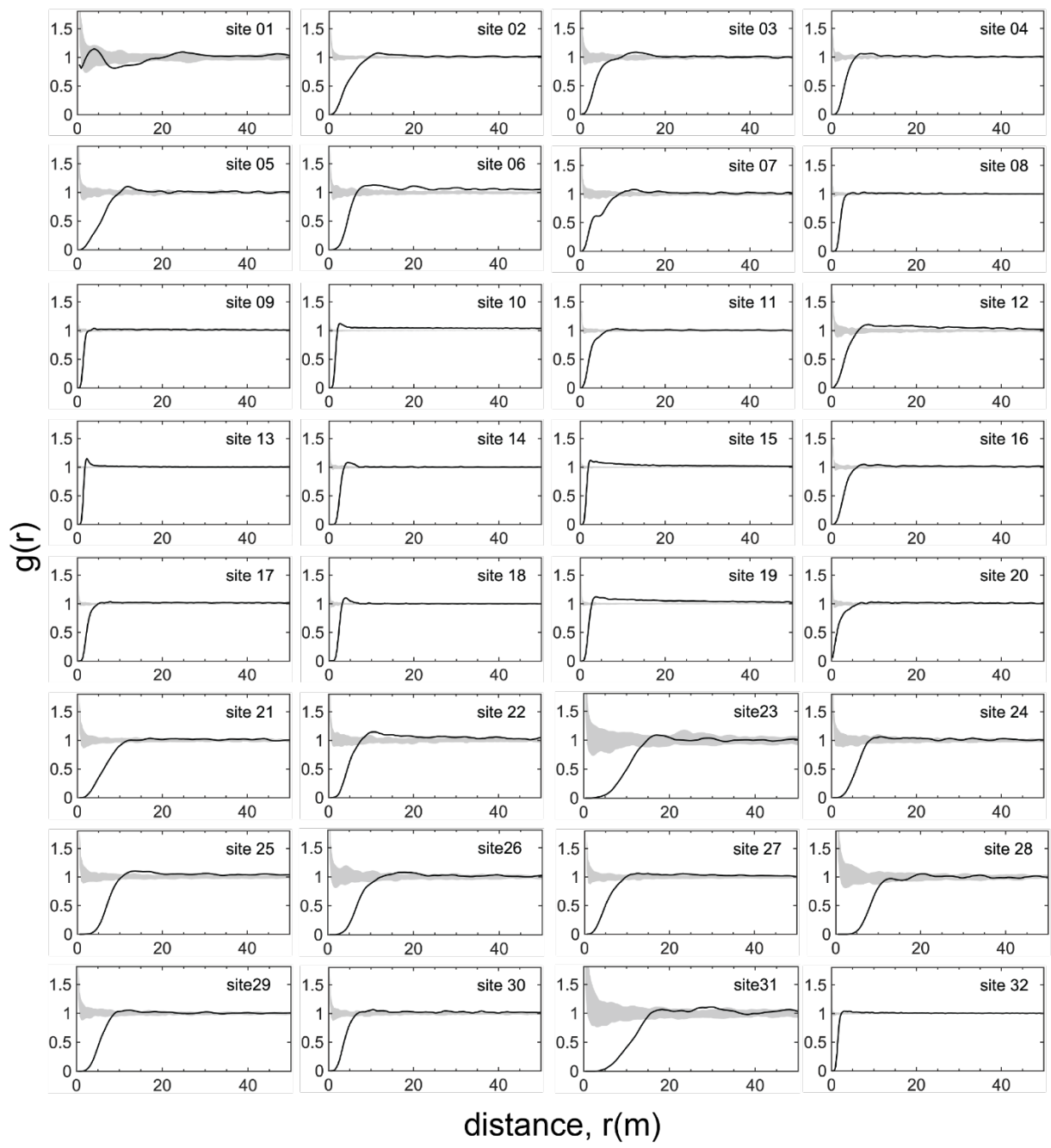

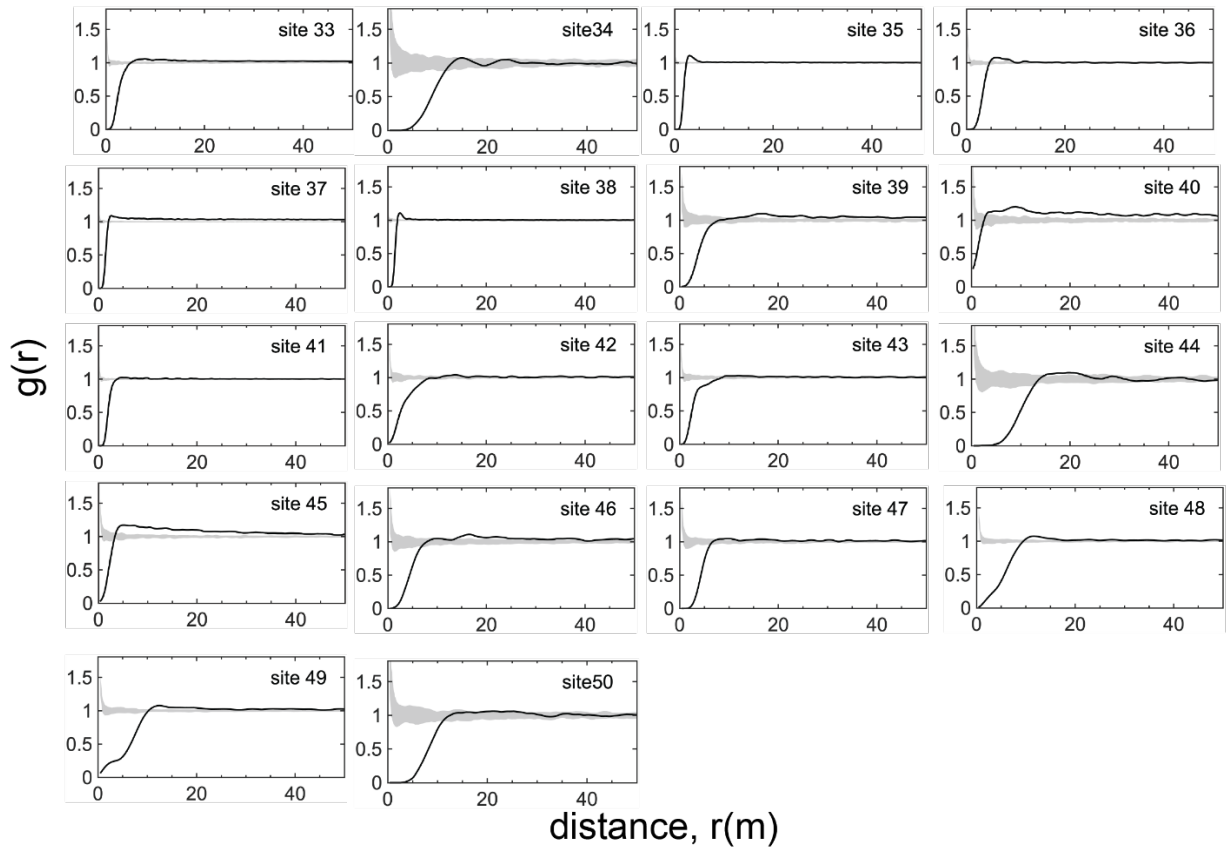

**Figure S1.** Results from the pair correlation *g*(*r*) function analysis for the 50 study sites. The gray bands represent 95% confidence intervals of complete spatial randomness (CSR). All study sites present 'small-scale over-dispersion and large-scale randomness'; some sites also present a signal of under-dispersion at intermediate scales.

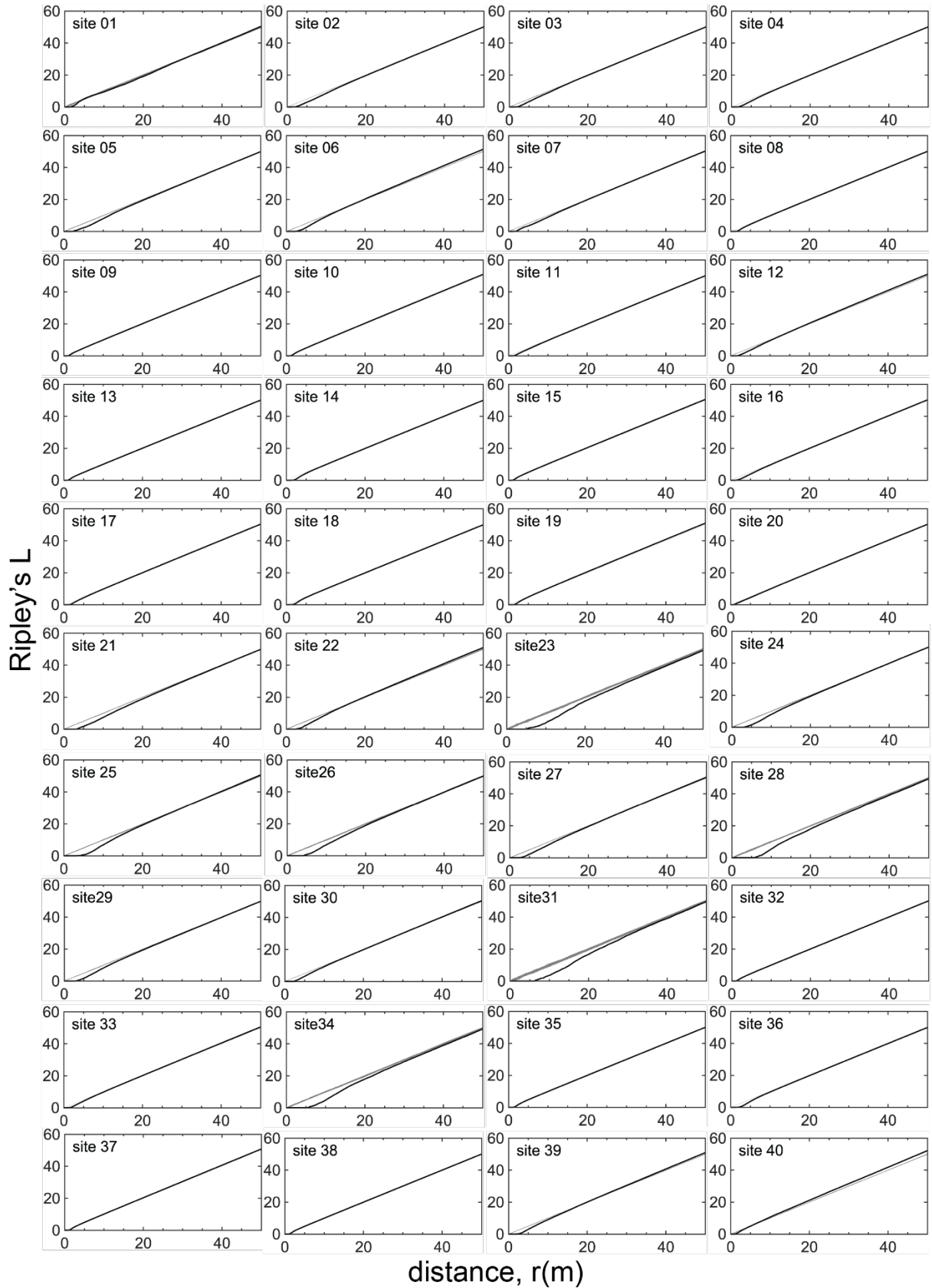

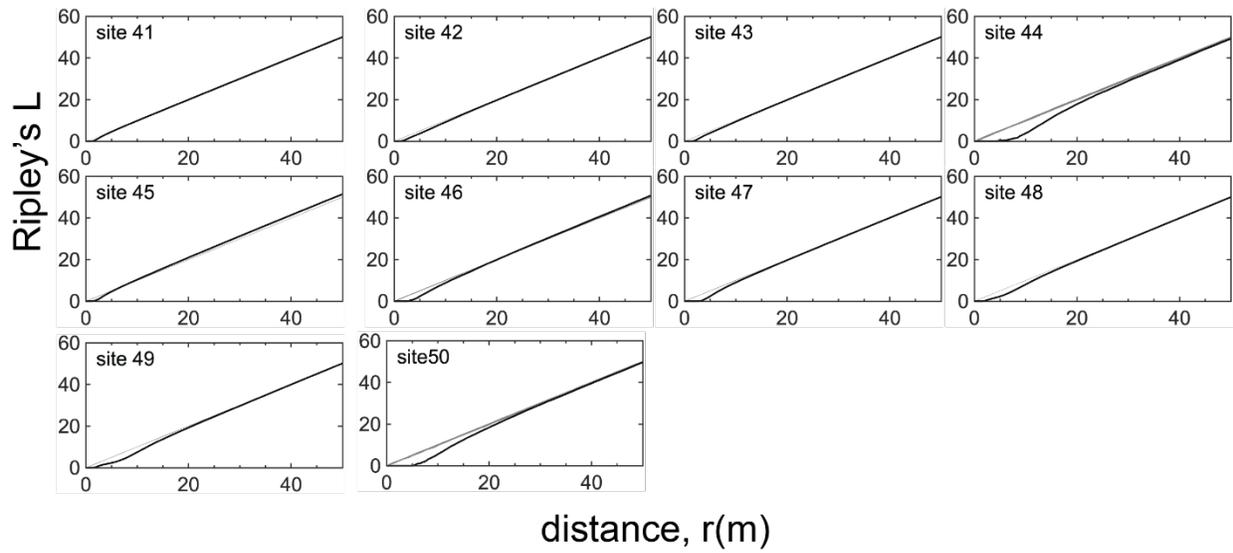

**Figure S2.** Results from the Ripley's *L*-function analysis for the 50 study sites. The gray bands represent 95% confidence intervals of complete spatial randomness (CSR).

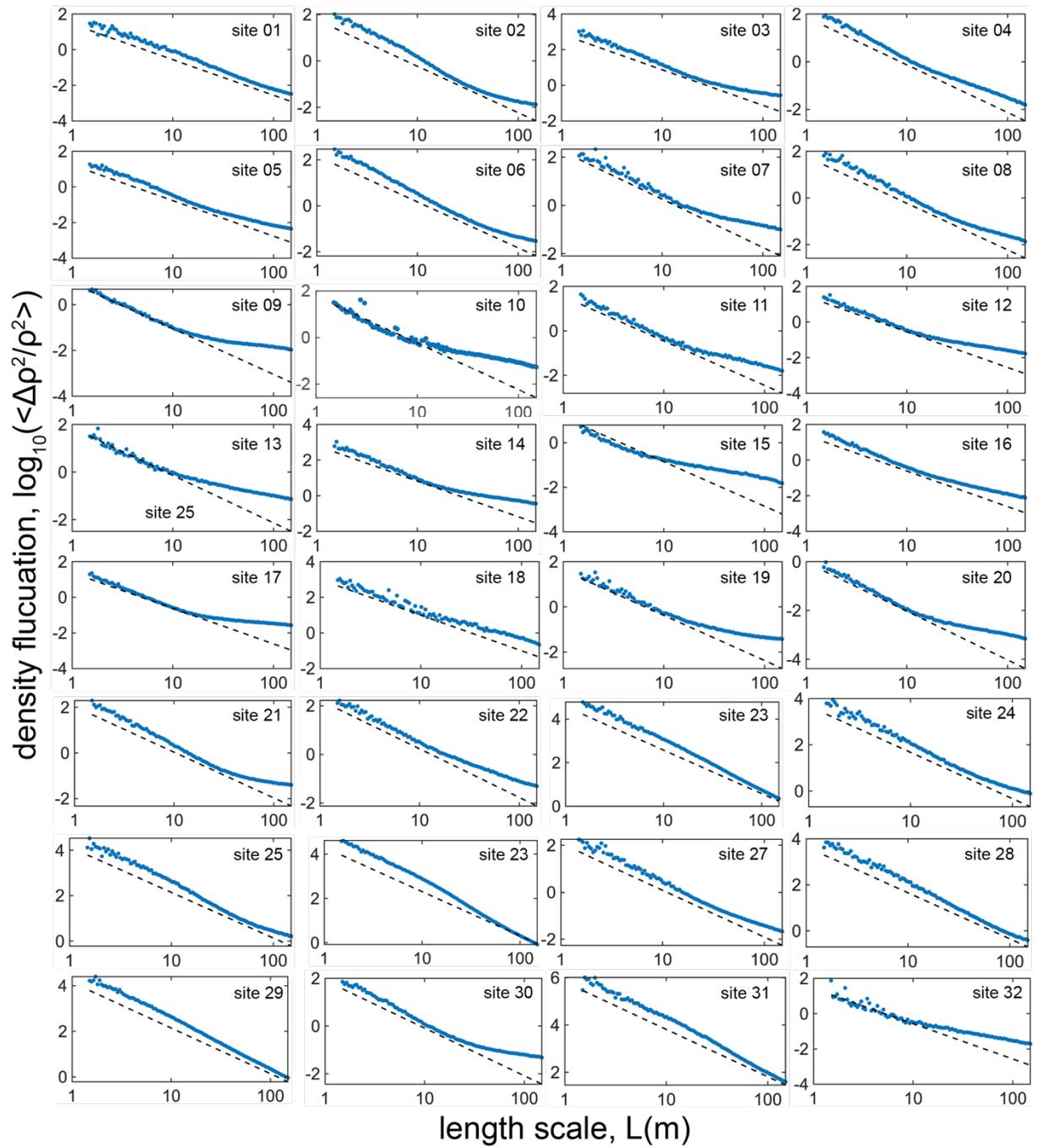

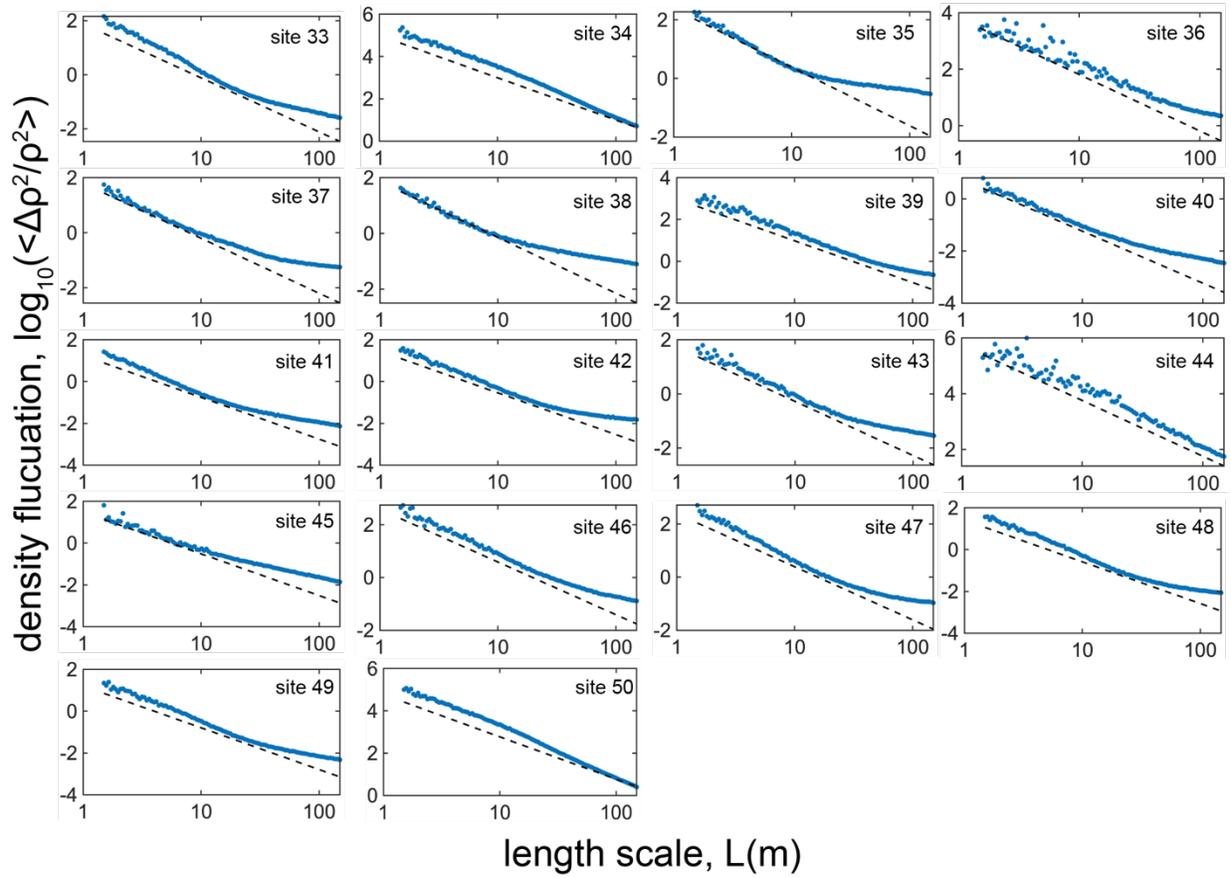

**Figure S3.** Results from the density fluctuations analysis for the 50 study sites. The dashed lines mark λ=2.0.

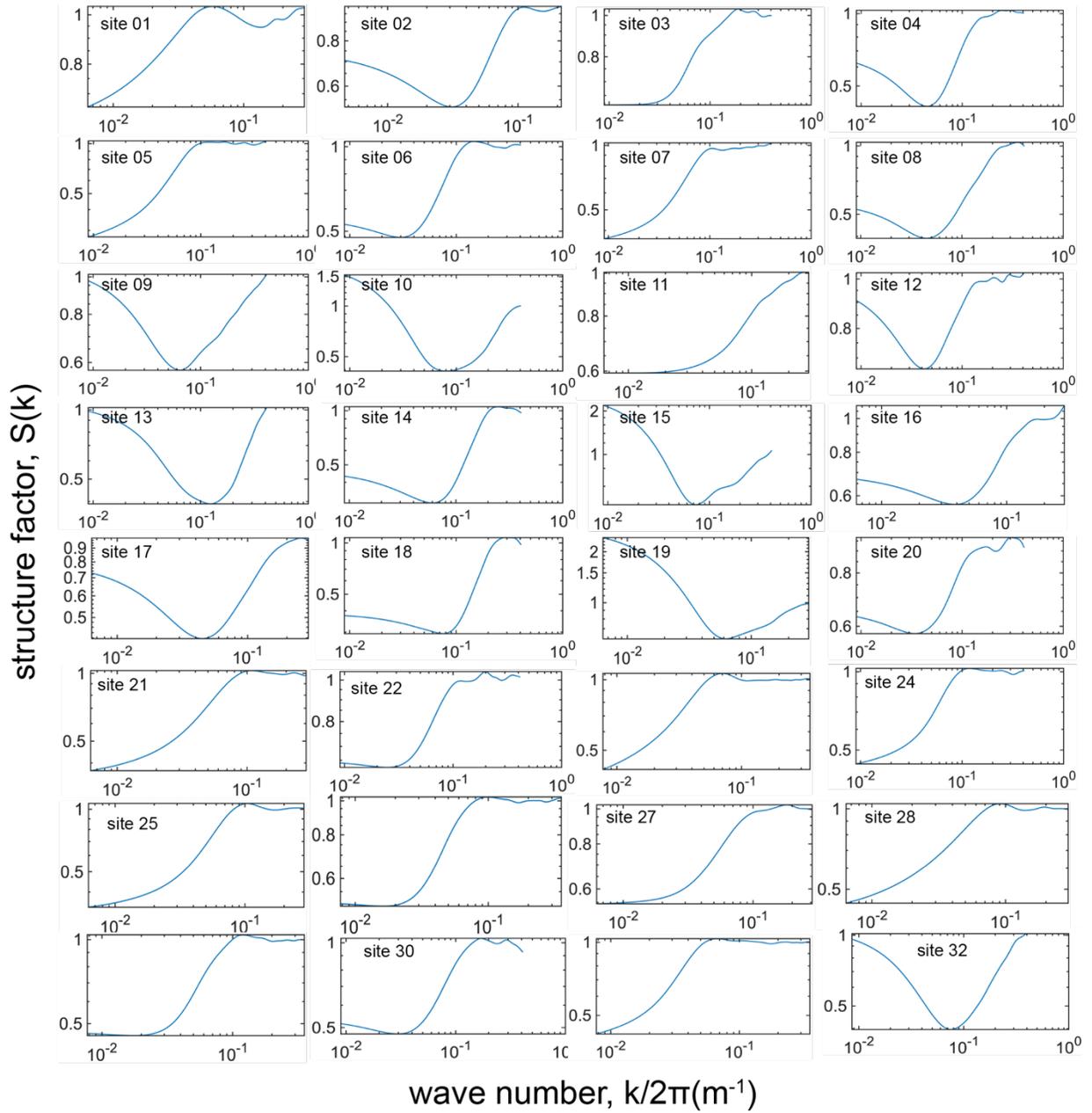

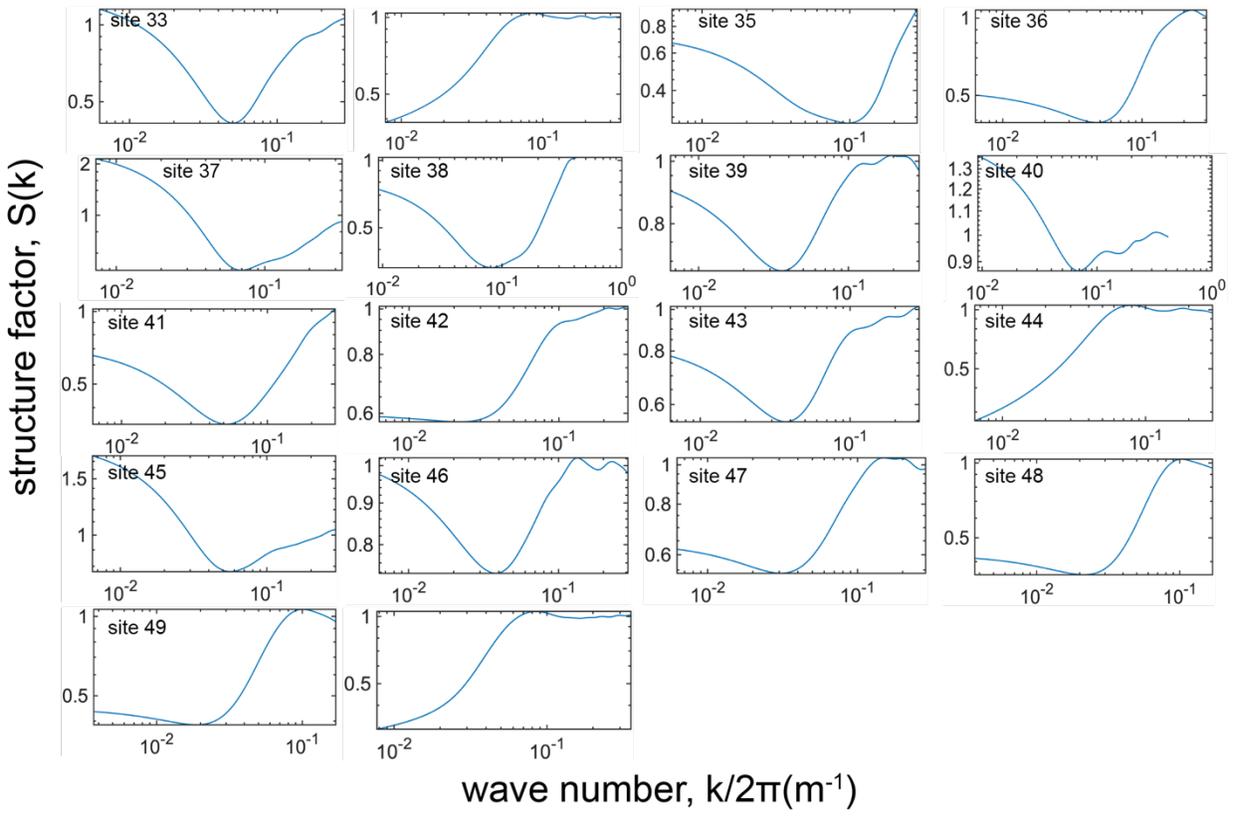

**Figure S4**. Results from the structure factor analysis for the 50 study sites.

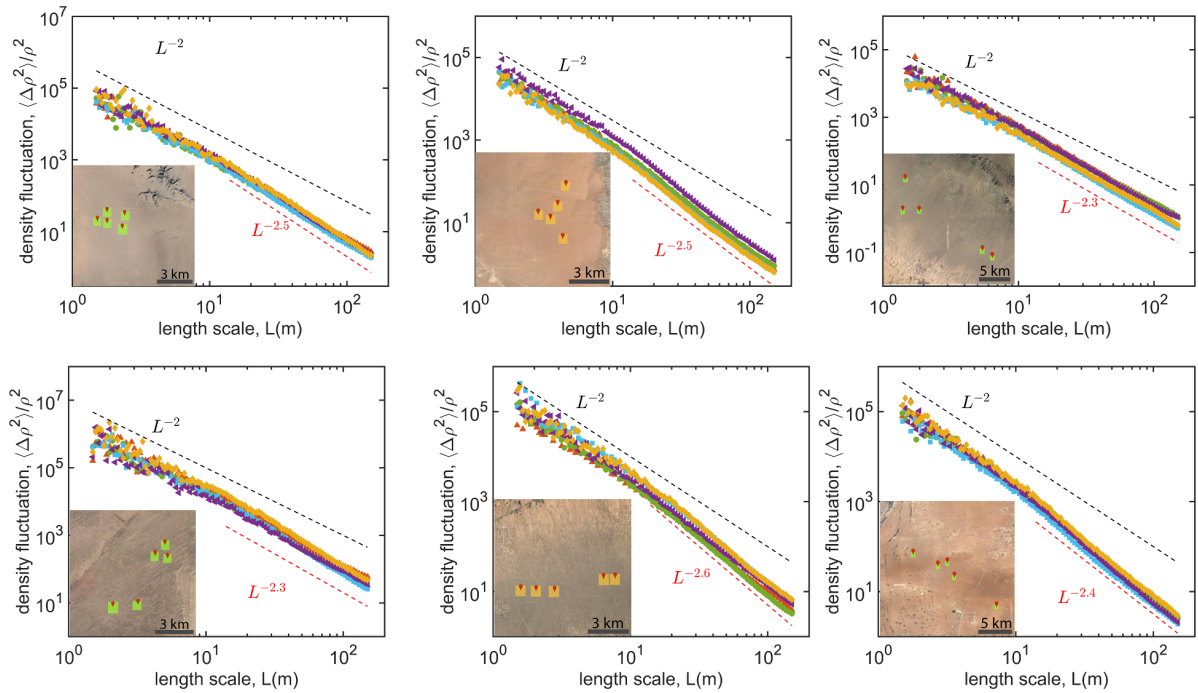

**Figure S5**. Density fluctuations analysis on the four additional sampling plots next to the study plots presenting signals of disordered hyperuniformity. The four additional plots were selected arbitrarily in proximity to the study plot. The insets illustrate the locations of the five plots altogether in each study site. The exponents of density fluctuations have values >2.0 for all additional plots, indicating that the signals of disordered hyperuniformity are robust across these studied landscapes.

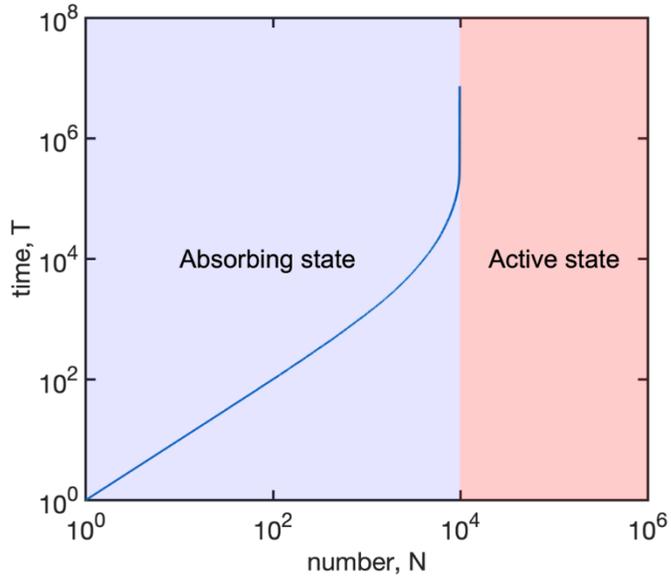

**Figure S6.** An illustration of the phase transition with increasing plant density in the ecological legacy model. Starting from a random initial state, the time steps needed for the system to reach an absorbing state (purple) increase linearly with $N$ when $N$ is below ~13370; however, when N crosses the critical density point > ~13370, the system transitions to an active state characterized by never-ending dynamics (pink).

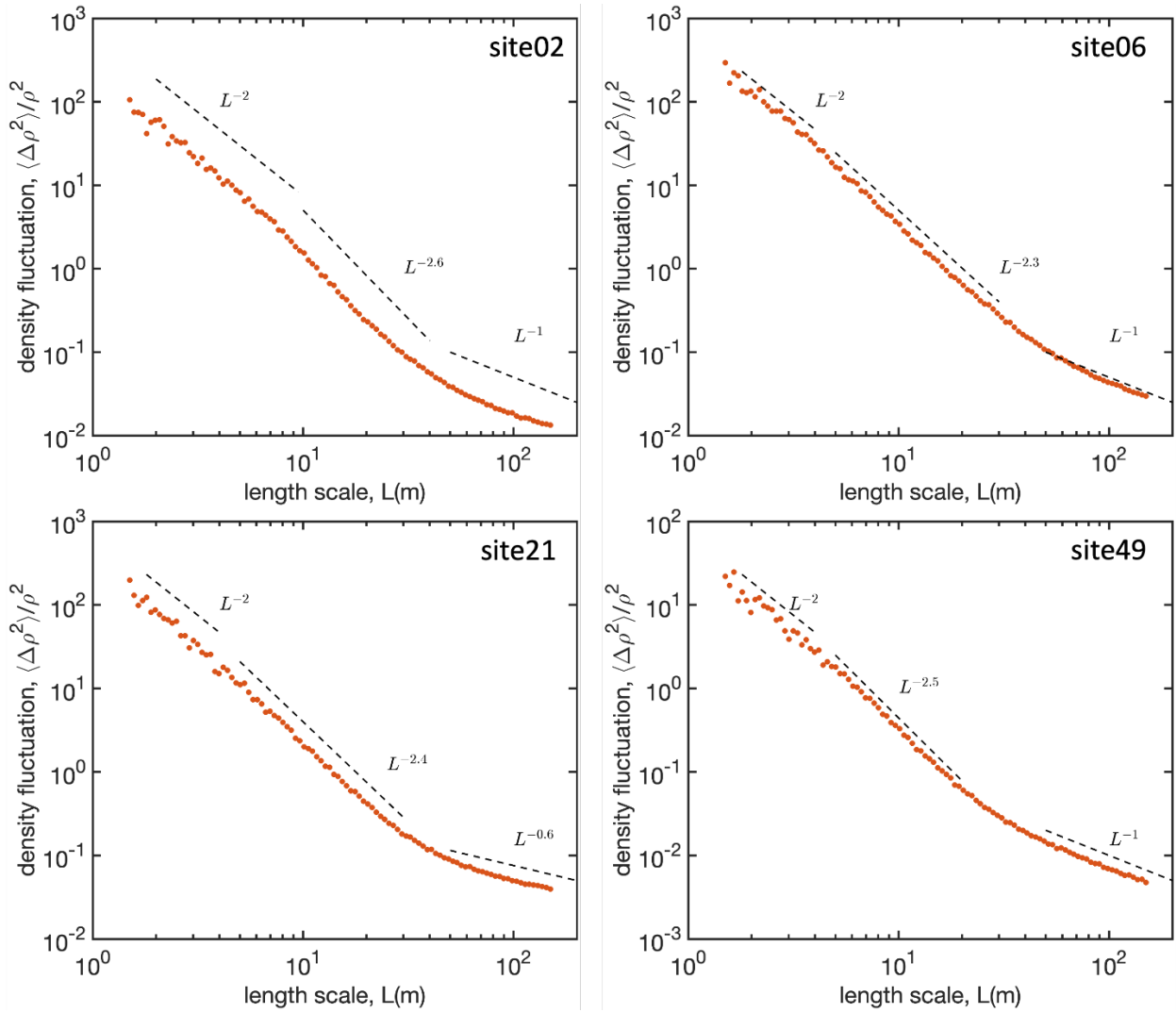

**Figure S7.** Four typical study sites presenting signals of hyperuniformity at intermediate scales. Among the 50 study sites, 22 exhibited density fluctuations (like the sites shown here) with exponent λ= 2.0 at small scales, λ>2.0 at intermediate scales (typically 10-30 m), and λ≤ 2.0 at larger scales (typically >30 m).

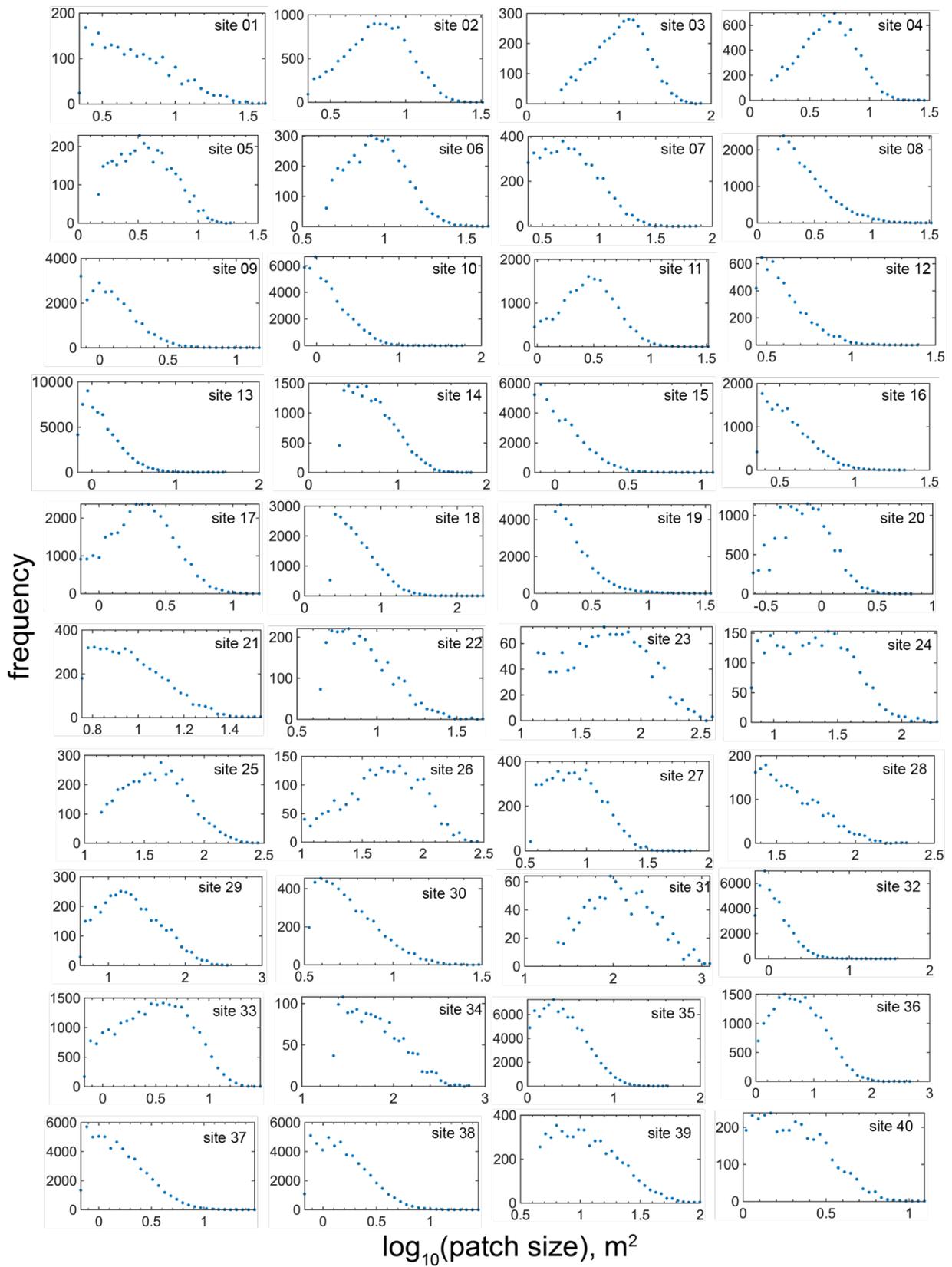

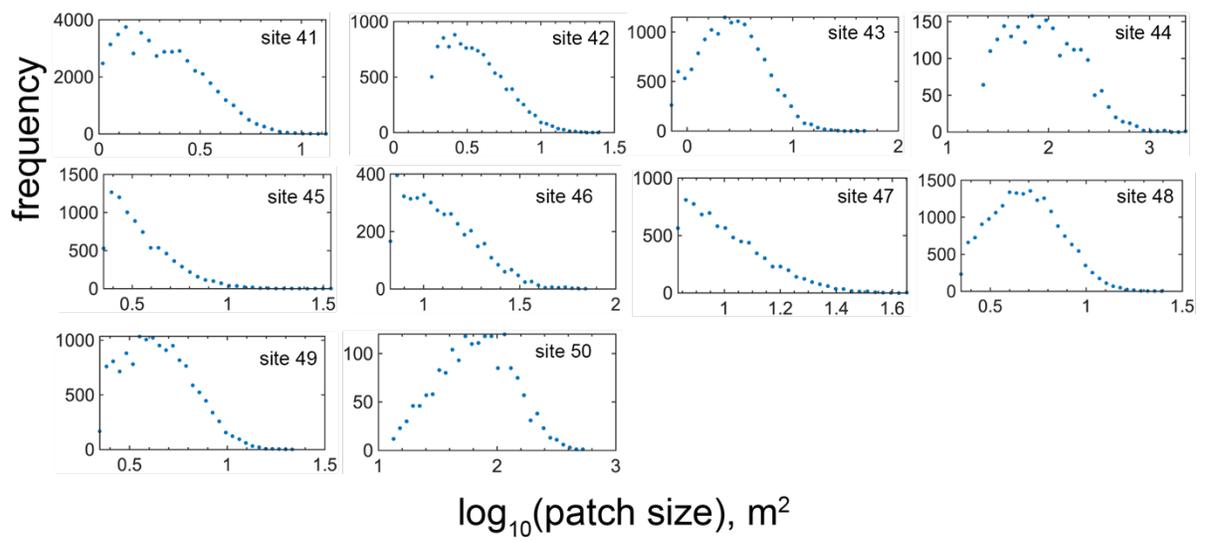

**Figure S8.** Vegetation patch-size distributions of all 50 study sites.

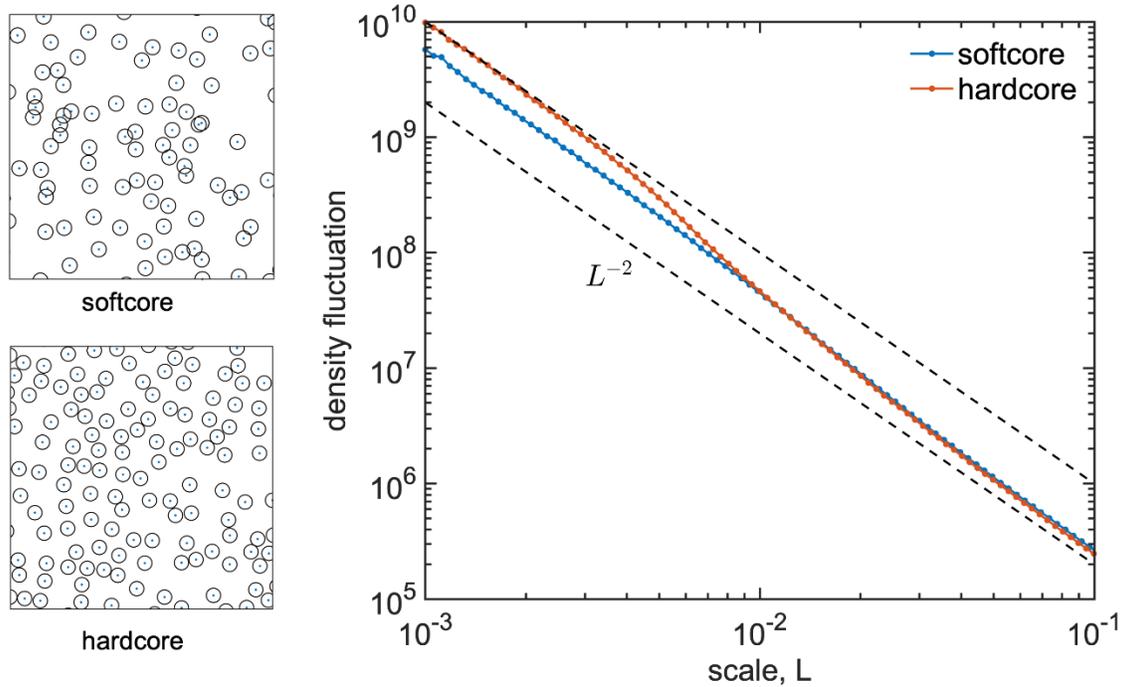

**Figure S9.** Comparison between the softcore and hardcore versions of the competition model. The competition model presents a robust signal of disordered hyperuniformity characterized by density fluctuations with exponent λ=2.0 at large spatial scales. See Methods for model descriptions.

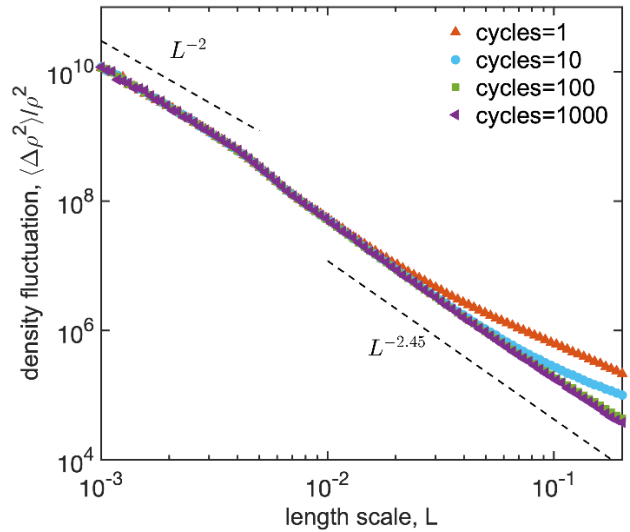

**Figure S10.** The ecological legacy model gives rise to signals of disordered hyperuniform patterns due to the random re-organization behavior at densities lower than the critical density point of phase transition (from the absorbing state to the active state). Upon the absorbing state (all re-generated individuals can survive) with 12 000 individuals (lower than the critical density with $\rho_c \approx 13370$), the system approaches disordered hyperuniformity (density fluctuations exponent $\lambda$ around 2.45) due to the random re-organization behavior (cycles 1-1000 are shown here). See Methods for model descriptions and reference (18) showing disordered hyperuniformity induced by random re-organization behaviors.

**Table S1.** Information on the 50 study sites.

| # | Longitude | Latitude | Aridity index | Density fluctuation exponent λ at long range | Hyperuniformity signal (λ>2.0) |
|---|---|---|---|---|---|
| 01 | -114.02101 | 31.98829 | 0.05 | 1.92 | Intermediate scale |
| 02 | -114.72889 | 32.35549 | 0.06 | 1.17 | Intermediate scale |
| 03 | -115.17788 | 32.80665 | 0.07 | 1.24 | Intermediate scale |
| 04 | -113.56134 | 32.83009 | 0.07 | 1.52 | Intermediate scale |
| 05 | -115.60855 | 32.11179 | 0.15 | 1.43 | Intermediate scale |
| 06 | -114.21212 | 33.89657 | 0.06 | 1.61 | Intermediate scale |
| 07 | -116.66338 | 35.00295 | 0.11 | 0.97 | None |
| 08 | -111.81889 | 36.67612 | 0.32 | 1.45 | Intermediate scale |
| 09 | -109.81577 | 36.76222 | 0.16 | 0.62 | None |
| 10 | -118.01995 | 40.51061 | 0.29 | 0.85 | None |
| 11 | -114.39858 | 36.68460 | 0.20 | 1.13 | None |
| 12 | -110.56953 | 38.40529 | 0.15 | 0.96 | None |
| 13 | -114.23884 | 41.20308 | 0.31 | 0.87 | None |
| 14 | -114.46514 | 39.39445 | 0.32 | 0.95 | None |
| 15 | -116.20336 | 41.92257 | 0.47 | 0.84 | None |
| 16 | -114.81928 | 35.30026 | 0.13 | 1.33 | None |
| 17 | -116.39976 | 36.65709 | 0.13 | 0.52 | None |
| 18 | -113.34151 | 38.13872 | 0.33 | 1.50 | None |
| 19 | -110.70874 | 35.43320 | 0.15 | 0.75 | Intermediate scale |
| 20 | -117.42707 | 36.33606 | 0.14 | 0.83 | None |
| 21 | -113.24461 | 31.51741 | 0.07 | 1.01 | Intermediate scale |
| 22 | -113.01269 | 31.13754 | 0.05 | 1.45 | None |
| 23 | -112.57025 | 30.29371 | 0.06 | 2.36 | Long range |
| 24 | -111.10379 | 30.21920 | 0.20 | 1.73 | Intermediate scale |
| 25 | -108.86041 | 32.44027 | 0.18 | 1.78 | Intermediate scale |
| 26 | -106.73753 | 31.97951 | 0.14 | 2.58 | Long range |
| 27 | -114.83697 | 30.86921 | 0.08 | 1.69 | Intermediate scale |
| 28 | -111.98486 | 29.00083 | 0.06 | 2.01 | Intermediate scale |
| 29 | -111.85867 | 29.65948 | 0.11 | 2.27 | Long range |
| 30 | -108.59346 | 36.28881 | 0.27 | 0.95 | None |
| 31 | -109.30290 | 34.79318 | 0.20 | 2.41 | Long range |
| 32 | -107.92553 | 42.03571 | 0.37 | 1.02 | None |
| 33 | -113.88044 | 27.37512 | 0.02 | 1.14 | Intermediate scale |
| 34 | -111.15562 | 32.06762 | 0.13 | 2.47 | Long range |
| 35 | -106.74168 | 34.94188 | 0.14 | 0.57 | None |
| 36 | -103.57476 | 26.93796 | 0.17 | 1.29 | None |
| 37 | -109.77088 | 41.76452 | 0.33 | 0.79 | None |
| 38 | -108.59260 | 45.44300 | 0.48 | 0.77 | None |
| 39 | -104.00758 | 33.42467 | 0.19 | 1.61 | Intermediate scale |
| 40 | -103.86831 | 35.78680 | 0.30 | 1.09 | None |
| 41 | -118.71704 | 38.94375 | 0.20 | 1.04 | None |
| 42 | -118.08833 | 35.13657 | 0.24 | 0.89 | Intermediate scale |
| 43 | -102.98840 | 25.45498 | 0.14 | 0.93 | Intermediate scale |
| 44 | -105.82696 | 29.38489 | 0.15 | 2.23 | Intermediate scale |
| 45 | -102.32015 | 32.28477 | 0.25 | 1.23 | None |
| 46 | -103.51506 | 32.03354 | 0.18 | 1.38 | Intermediate scale |
| 47 | -115.80582 | 33.76243 | 0.12 | 1.17 | Intermediate scale |
| 48 | -114.79326 | 32.34671 | 0.06 | 0.92 | Intermediate scale |
| 49 | -114.67807 | 32.19605 | 0.07 | 1.06 | Intermediate scale |
| 50 | -106.22024 | 32.08496 | 0.14 | 2.52 | Long range |

**Table S2.** A brief description of the model parameters and state variables.

| Model | Symbol | Description | Value Range | Default Values |
|---|---|---|---|---|
| **Competition model** | $R_c$ | Interaction range (competition) | 0-0.1 | 0.006 (Fig. 3C-E, Fig. 5, Fig. 6B); 0.006-0.01 (Fig. 3F) |
| **Ecological legacy model** | $R_c$ | Interaction range (competition; facilitation) | 0-0.1 | 0.006 (Fig. 3C-E, Fig. 5, Fig. 6B); 0.006-0.01 (Fig. 3F) |
| | $M_r$ | Random mortality rate | 0-1 | 0.2 |
| **Scale-dependent feedbacks model** | $M_i$ | Mortality probability of individual $i$ | 0-1 | - |
| | $h_e$ | The harshness of the environment | 0-1 | 0.05, 0.15, 0.35, 0.5, 0.65 (Fig. 3); 0.05 (Fig. 6A). |
| | $d_j$ | Distance from $i$ to $j$ | 0-10 | - |
| | $p_f$ | Exponent in the Hill function for facilitation | 4 | 4 |
| | $p_c$ | Exponent in the Hill function for competition | 2 | 2 |
| | $h_c$ | Attenuation coefficient of competitive effect | 1 | 1 |
| | $h_f$ | Attenuation coefficient of facilitation effect | 1.5 | 1.5 |
| | $C_\alpha$ | Competition intensity coefficient | 0.88 | 0.88; 0.09 (scale-dependent feedback model in Fig. 6A) |
| | $D$ | Individual diameter | 0.55 | 0.55 |
| **Density-dependent aggregation model** | $S_s$ | Concentration of suspended/aeolian sediment | \ | A matrix with an initial uniform distribution ranging from 0 to 1. |
| | $S_d$ | The height of plant trapped sediment | \ | A matrix with an initial uniform distribution ranging from 0 to 1. |
| | $D_s$ | Diffusion constant of $S_s$ | 1 | 1 |
| | $D_d$ | Diffusion constant of $S_d$ | 0.01 | 0.01 |
| | $\alpha, \beta$ | Erosion constants | 1 | 1 |
| | $\varepsilon$ | Correlation constant of $D(P)$ and $P(S_d)$ | 1 | 1 |
| | $\xi$ | Correlation constant of $P(S_d)$ and $S_d$ | 1 | 1 |
| | $\tau$ | The rate of $S_d$ into $S_s$ | 3-7 | 3, 4, 5, 6, 7 |